\documentclass[mnsc, nonblindrev]{IISEnew} % current default for manuscript submission

\OneAndAHalfSpacedXI % current default line spacing
\usepackage{amsmath}
\usepackage{rotating}
\usepackage{xcolor}
\usepackage{multirow}
\usepackage{caption}
\usepackage{amsfonts}
\usepackage{blkarray}
\usepackage{float}%To place figure at required positions.
\usepackage{caption} 
\usepackage{rotating}
\usepackage{comment}
\usepackage[hidelinks]{hyperref}% package for hyperlinks without ugly borders.
\usepackage{multirow}
\usepackage{hhline}
\usepackage{array}
\newcolumntype{L}[1]{>{\raggedright\let\newline\\\arraybackslash\hspace{0pt}}m{#1}}
\newcolumntype{C}[1]{>{\centering\let\newline\\\arraybackslash\hspace{0pt}}m{#1}}
\newcolumntype{R}[1]{>{\raggedleft\let\newline\\\arraybackslash\hspace{0pt}}m{#1}}
\usepackage{color}
\usepackage{array}
\usepackage{mathtools}
\usepackage{mleftright}
\usepackage[utf8]{inputenc}
\newcommand\numberthis{\addtocounter{equation}{1}\tag{\theequation}}
\usepackage{enumerate}

\usepackage[utf8]{inputenc}
\usepackage{epstopdf}
\usepackage[title]{appendix}
\usepackage{epstopdf}
\usepackage{lscape}
\usepackage{epsfig}
\usepackage{caption}
\TheoremsNumberedThrough     % Preferred (Theorem 1, Lemma 1, Theorem 2)
\ECRepeatTheorems

\EquationsNumberedThrough    % Default: (1), (2), ...
\begin{document}
%%%%%%%%%%%%%%%%

%\setcounter{page}{1}
%\DOI{10.1287/xxxx.0000.0000}%

%\RUNAUTHOR{Suman and Krishnamurthy}

%\RUNTITLE{Analysis of Two-Station Polling Queues with Setups}

% \TITLE{Bundling Information Goods of Decreasing Value}
\TITLE{Analysis of Two-Station Polling Queues with Setups using Continuous Time Markov Chain}

\ARTICLEAUTHORS{%
\AUTHOR{Ravi Suman}
\AFF{Department of Industrial and Systems Engineering, University of Wisconsin-Madison, Madison, WI 53705, USA \EMAIL{rsuman@wisc.edu}} %, \URL{}}
\AUTHOR{Ananth Krishnamurthy}
\AFF{Decision Sciences, Indian Institute of Management Bangalore, Bangalore 560076, India \EMAIL{ananthk@iimb.ac.in}}
} % end of the block

\ABSTRACT{%
The paper analyzes the performance of tandem network of polling queue with setups. For a system with two-products and two-stations, we propose a new approach based on a partially-collapsible state-space characterization to reduce state-space complexity. In this approach, the size of the state-space is varied depending on the information needed to determine buffer levels and waiting times. We evaluate system performance under different system setting and comment on the numerical accuracy of the approach as well as provide managerial insights. Numerical results show that approach yields reliable estimates of the performance measures. We also show how product and station asymmetry significantly affect the systems performance.
% Enter your abstract
}%

%\KEYWORDS{deterministic inventory theory; infinite linear programming duality;
%  existence of optimal policies; semi-Markov decision process; cyclic schedule}


\maketitle
%%%%%%%%%%%%%%%%%%%%%%%%%%%%%%%%%%%%%%%%%%%%%%%%%%%%%%%%%%%%%%%%%%%%%%

% Samples of sectioning (and labeling) in MNSC
% NOTE: (1) \section and \subsection do NOT end with a period
%       (2) \subsubsection and lower need end punctuation
%       (3) capitalization is as shown (title style).
%
%\section{Introduction.}\label{intro} %%1.
%\subsection{Duality and the Classical EOQ Problem.}\label{class-EOQ} %% 1.1.
%\subsection{Outline.}\label{outline1} %% 1.2.
%\subsubsection{Cyclic Schedules for the General Deterministic SMDP.}
%  \label{cyclic-schedules} %% 1.2.1
%\section{Problem Description.}\label{problemdescription} %% 2.

% Text of your paper here

\section{Introduction}\label{Introduction6}
Polling queues find applications when multiple products compete for a common resource. In a polling queue, a single server serves multiple queues of products, visiting the queues one at a time in a fixed cyclic manner. In manufacturing, polling queues have been used to model flow of multiple products undergoing manufacturing operations in a factory. In healthcare, polling queues have been used to model the flow of different types of patients through various activities in a hospital or clinic. In transportation, polling queues have been used to model multiple traffic flows in a transportation network. Comprehensive survey on the analysis of polling queues can be found in $\left(\text{Takagi }\cite{Takagi2000}, \text{Vishnevskii \& Semenova }\cite{Vishnevskii2006}\right)$.\\

While a majority of existing research on polling queues focus on the single-station polling queue, this work focuses on the analysis of a tandem network of polling queues with setups. Our motivation for studying tandem network of polling queues with setups is derived from our collaboration with a large manufacturer of rolled aluminum products $\left(\text{RAP}\right)$ where the manufacturing operations can be modeled as a tandem network of polling queues. At this facility, the manufacturing process involves steps like rolling of aluminum ingots into plates, heat treating to improve properties, stretching the plates to improve straightness, aging to cure the metal, sawing the plates into smaller pieces, and conducting ultrasonic inspection to check material properties. In this case, each manufacturing operation can be modeled as a polling queue, processing different types of alloys, and incurring a setup when the equipment switches from one type of product to another type of product in a sequential manner. A particular product may be processed through a series of these operations based on either a predetermined or probabilistic sequence of operations. In such a setting, estimates of mean waiting time can help managers release and schedule jobs, quote lead times for customers, and improve coordination with downstream operations.\\

Tandem network of polling queues also find application in factories of process/semi-process industries such as chemical, plastic, and food industries where significant setup times are incurred when a machine switches from producing one type of product to another. To reduce cost, manufacturers often produce their products in batches, and use an exhaustive policy, i.e, serve all products waiting in a queue before switching over to another product type. Thus, determining the impact of setup times on waiting times is of key interest to the managers.\\

Despite the importance of tandem network of polling queues, there has been limited studies of such networks. Exact analysis of polling models is only possible in some cases, and even then numerical techniques are usually required to obtain waiting times at each queue. We propose a decomposition based approach for the analysis of the performance of tandem network of polling models. Our research makes two key contributions. First, we provide a computationally efficient method that exploits the structure of the state-space to provide solutions for tandem polling queues with setups. In particular, we use a partially-collapsible state-space approach that captures or ignores queue length information as needed in the analysis. We show that this approach reduces computational complexity and provides reasonable accuracy in performance estimation. Second, we investigate the impact of different manufacturing settings, such as, location of bottleneck stations, asymmetry in waiting times, and setup times on systems performance measures. We find that the location of bottleneck station and differences in service rates can have significant impact on the waiting times.\\

The rest of the paper is organized as follows. In Section \ref{LiteratureReview6}, we provide a brief literature review on polling queues and analysis of tandem network of queues. We describe the system in Section \ref{SystemDescription6} and the approach used to analyze the two-station system in Section \ref{Subsystem1} and Section \ref{Subsystem2}. In Section \ref{NumericalResults6}, we validate our approach and provide useful numerical insights. Finally, we conclude and provide future extensions in Section \ref{Conclusions6}.
\section{Literature Review}\label{LiteratureReview6}
Polling queues and their applications have been an active field of research for the past few decades. Takagi \cite{Takagi2000}, Vishnevskii  and Semenova \cite{Vishnevskii2006}, and  Boona et al. \cite{Boona11} provide a comprehensive survey on polling queues and their applications. We group our discussion of the literature in three categories$\colon$ polling queue with zero setups, polling queue with non-zero setups, and network of polling queues.\\

\textbf{Polling queue with zero setups}$\colon$ One of the earliest techniques for analyzing polling queues with zero setups uses a \emph{server vacation model}, where the server periodically leaves a queue and takes a vacation to serve other queues. Fuhrmann et al. \cite{Fuhrmann85} uses such a vacation model to study a symmetric polling station with $Q$ queues served in a cyclic order by a single server and determines the expressions for sojourn times under exhaustive, gated, and $k$-limited service discipline. They show that the stationary number of customers in a single station polling queue (summed over all the queues) can be written as the sum of three independent random variables$\colon\left(i\right)$ the stationary number of customers in a standard M/G/I queue with a dedicated server, $\left(ii\right)$ the number of customers in the system when the server begins an arbitrary vacation (changeover), and $\left(iii\right)$ number of arrivals in the system during the changeover. Boxma et al. \cite{Boxma87} use a stochastic decomposition to estimate the amount of work (time needed to serve a specific number of customers) in cyclic-service systems with hybrid service strategies (e.g., semi-exhaustive for first product class, exhaustive for second and third product class, and gated for remaining product classes) and use the decomposition results to obtain a pseudo-conservation law for such cyclic systems.\\

\textbf{Polling queue with non-zero setups}$\colon$ Several studies have used transform methods to find the distributions for waiting times, cycle times, and queue lengths in a single-station polling queue with setups. Cooper et al. \cite{RBCooper96} propose a decomposition theorem for polling queues with non-zero switchover times and show that the mean waiting times is the sum of two terms$\colon\left(\text{1}\right)$ the mean waiting time in a "corresponding" model in which the switchover times are zero, and $\left(\text{2}\right)$ a simple term that is a function of mean switchover times. Srinivasan et al. \cite{Srinivasan95} use Laplace–Stieltjes Transform $\left(\text{LST}\right)$ methods to compute the moments of the waiting times in $R$ polling queues with nonzero-setup-times for exhaustive and gated service. The algorithm proposed requires estimation of parameters with $\log{\left(R\mathcal{E}\right)}$ complexity, with $\mathcal{E}$ as the desired level of accuracy. Once the parameters have been calculated, mean waiting times may be computed with $\mathcal{O}\left(R\right)$ elementary operations. Borst and Boxma \cite{Borst97} generalize the approach used by Srinivasan et al. \cite{Srinivasan95} to derive the joint queue length distribution for any service policy. Boxma et al. \cite{Boxma09} analyzes a polling system of $R$-queues with setup times operating under gated policy and determine the LST for cycle times under different scheduling disciplines such as FIFO and LIFO. They show that LST of cycle times is only dependent on the polling discipline at each queue and is independent of the scheduling discipline used within each queue.\\

In addition to LST techniques, mean value analysis has also been used to estimate performance measures for polling queues with nonzero setups. Hirayama et al. \cite{Hirayama04} developed a method for obtaining the mean waiting times conditioned on the state of the system at an arrival epoch. Using this analysis, they obtain a set of linear functional equations for the conditional waiting times. By applying a limiting procedure, they derive a set of $R(R +1)$ linear equations for the unconditional mean waiting times, which can be solved in $\mathcal{O}\left(R^6\right)$ operations. Winands et al. \cite{Winands06} calculates the mean waiting times in a single-station multi-class polling queue with setups for both exhaustive and gated service disciplines. They use mean value analysis to determine the mean waiting times at the polling queue. They derive a set of $R^{2}$ and $R\left(R +1\right)$ linear equations for waiting time figures in case of exhaustive and gated service. In these studies of polling queues using LST techniques or mean value analysis, the authors have restricted their scope of study to single-station polling queues. Extending their approach to tandem network of polling queue will increase the computational complexity. Therefore, in our work, we propose a decomposition based approach.\\

$\textbf{Network of polling queues}\colon$ Altman and Yechiali \cite{Altman94} study a closed queueing network for token ring protocols with $Q$ polling stations, where a product upon completion of the service is routed to another queue probabilistically. They determine explicit expressions for the probability generating function for the number of products at various queues. However, the system considered is closed system with $N$ products in circulation, which could be a restrictive assumption in some applications. Jennings \cite{Jennings08} conducts a heavy traffic analysis of two polling queues for two stations in series and prove limit theorems for exhaustive and gated discipline for the diffusion scaled, two-dimensional total workload process using heavy traffic analysis. Suman and Krishnamurthy (\cite{Suman18} -- \cite{Suman21}) study a two-product two-station tandem network of polling queues with finite buffers using Matrix-Geometric approach. However, the analysis is restricted to systems with small buffer capacity. In comparison, this paper analyzes an open network of two polling queues with exogenous arrivals using decomposition.\\
\section{System Description and Overview of Approach}\label{SystemDescription6}
In this section, we describe the system and provide an overview of the approach to estimate performance measures for the system.
\subsection{System Description}\label{System Description}
\begin{figure}[h!]
\begin{center}
\includegraphics[scale=0.32]{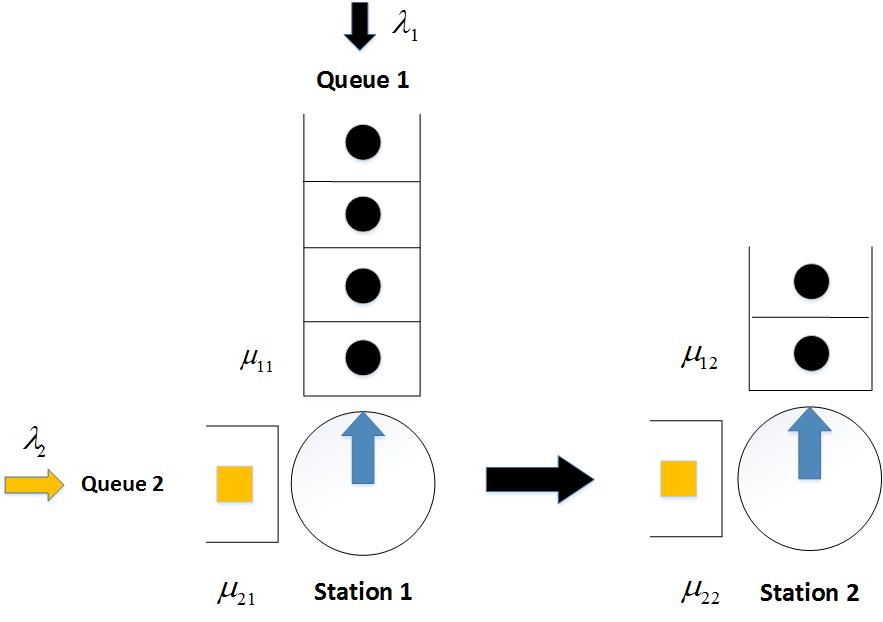}
\caption{Network of two-products two-stations polling queue.}
\label{fig:mesh6.1}
\end{center}
\end{figure}

We analyze a tandem polling queue with infinite capacity, each with two product types, indexed by $i$, for $i ={}1, 2$ operating under \emph{independent polling strategy}. Products of type $i$ arrive from the outside world to their respective queue at station 1 according to independent Poisson process with parameter $\lambda_i$. Each product type is served by a single server at station $j$, for $j ={}1, 2$ in a fixed cyclic manner $\left(\text{see Figure } \ref{fig:mesh6.1}\right)$ following an exhaustive service policy. Under the \emph{independent polling strategy}, at each station, the server switches to serve products of the other type after emptying the queue being served, independent of the state of the other station. After service at station 1, the product proceeds from station 1 to station 2, and exits the system after the service is completed at station 2. Service times at these stations for product $i$ has an exponential distribution with parameter $\mu_{ij}$ at station $j$. When a server switches from queue $i'$ to queue $i$, for $i' ={}1, 2$ and $i' \neq i$, at station $j$, the server incurs a setup time $H_{ij}$ that has an exponential distribution with rate $\mu_{s_{ij}}$. We assume that the setups are state independent, i.e., the server incurs a setup time at the polled queue whether or not products are waiting at the queue. We also assume that setup times are independent of service times and other queue type. Note that the system is stable when $\sum_{i={}1}^{2} \lambda_{i}\mu_{ij}^{-1}< $ 1 for each $j$. We assume this condition holds for our system. \\

\graphicspath {{Figures/}}

The goal is to calculate the following system performance measures$\colon$ (i) average buffer level, $\mathbb{E}\left[L_{ij}\right]$, defined as the average amount of material stored in buffer for product type $i$ at station $j \left[\text{parts}\right]$ and (ii) average waiting time, $\mathbb{E}\left[W_{i}\right]$, defined as the average time required by products to go through station 1 and 2 $\left[\text{time units}\right]$.\\

To solve the system described above using a conventional Markov chain $\left(\text{MC}\right)$ approach, we would need to use a six-tuple state space resulting in over 2.5 million states for a system with a buffer size of 20. To address this curse-of-dimensionality, we propose a new approach based on decomposition. We first describe the general approach and provide details in Section \ref{Subsystem1} and \ref{Subsystem2}.\\

\subsection{Overview of Approach}\label{Approach}
\begin{figure}[h!]
\begin{center}
\includegraphics[scale=0.70]{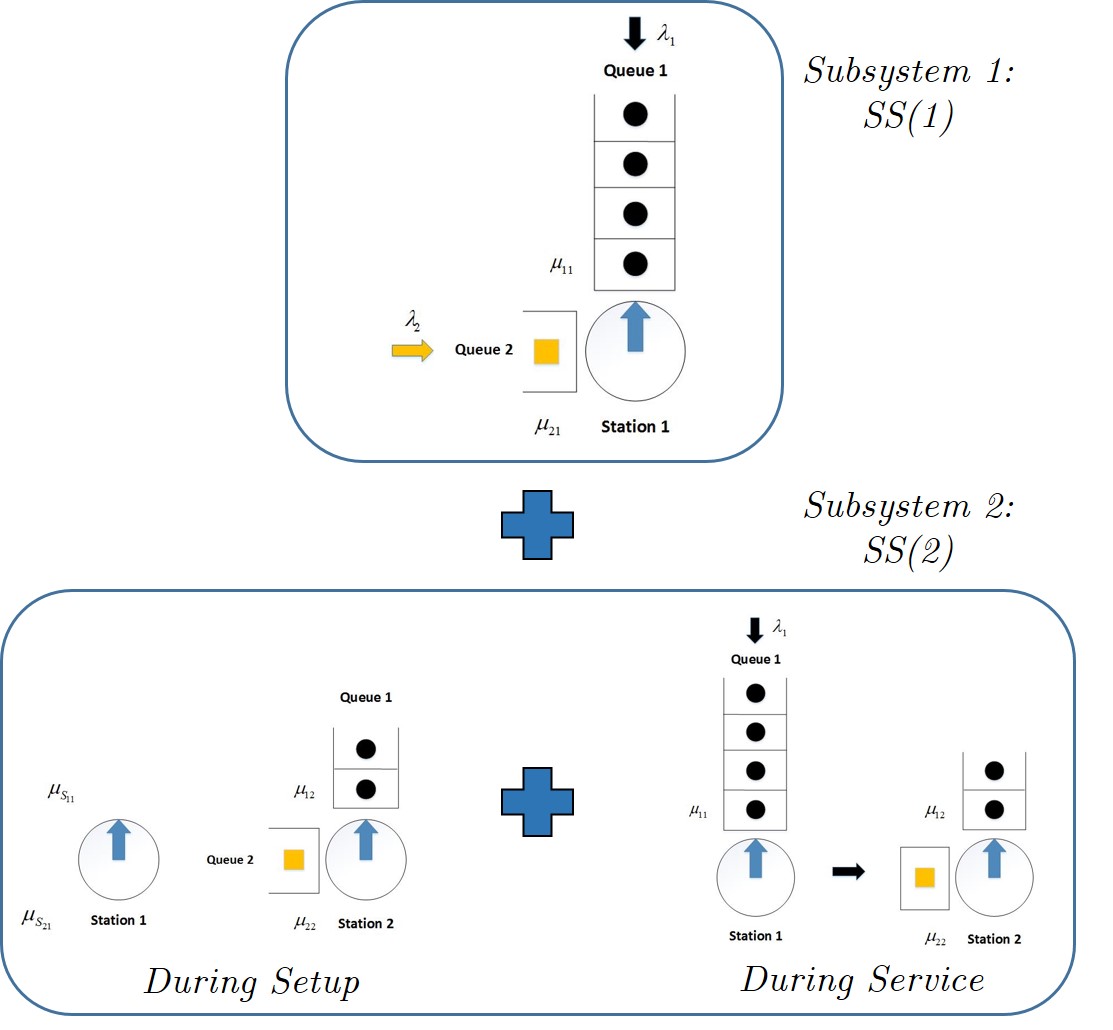}
\caption{Decomposition of system into subsystem $SS\left(1\right)$ and $SS\left(2\right)$.}
\label{fig:mesh6.2}
\end{center}
\end{figure}

The main idea is to decompose the two-station polling queue into two subsystems$\colon SS\left(k\right)$  for $k ={}1, 2$ as shown in Figure $\left(\ref{fig:mesh6.2}\right)$, and study each subsystem independently. Subsystem $SS\left(1\right)$ comprise of only station 1 of the system. We use exact analysis methods for subsystem $SS\left(1\right)$ to obtain performance measures at station 1. Subsystem $SS\left(2\right)$ comprise of both station 1 and station 2. We analyze subsystem $SS\left(2\right)$ to estimate performance measures at station 2. Since arrivals at station 2 depend on departures from station 1, the analysis of subsystem $SS\left(2\right)$ requires joint analysis of station 1 and station 2. In solving the subsystem $SS\left(2\right)$, we make use of the fact that the service policy adopted by the server is exhaustive at both the stations, and that the queue becomes zero for the served product type before it switches to serve another product. We exploit this fact to define the `partially-collapsible state-space' needed to analyze subsystem $SS\left(2\right)$. In this partially-collapsible state-space, the size of the state-space is varied depending on the information that needs to be retained to conduct the analysis. We use a combination of state-space description with four-tuples and five-tuples to model the relevant state transitions in subsystem $SS\left(2\right)$ depending on if the server at station 1 is doing a setup, or serving products, respectively. This approach helps reduce the state complexity and yet yields good approximations for the performance measures at station 2. The details are provided in the next section.
\graphicspath {{Figures/}}

\newpage

\section{Analysis of Subsystem $SS\left(1\right)$}\label{Subsystem1}
\begin{figure}[H]
\begin{center}
\includegraphics[scale=0.40]{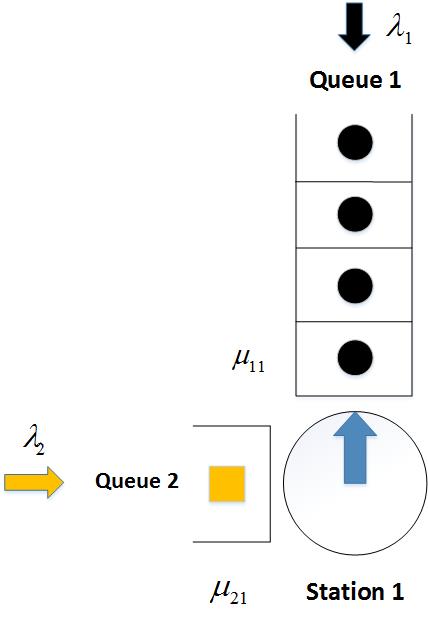}
\caption{Illustration of subsystem $SS\left(1\right)$.}
\label{fig:mesh6.3}
\end{center}
\end{figure}

In subsystem $SS\left(1\right)$, we consider only station 1 of the system described in Figure \ref{fig:mesh6.1}. In this, we consider system of single server serving two product types as shown in Figure \ref{fig:mesh6.3}. We analyze this subsystem to estimate performance measures for station 1. It should be noted that the  subsystem $SS\left(1\right)$ can be analyzed using mean value approach in Winands et al. \cite{Winands06} or using Laplacian approach in Boxma et al. \cite{Boxma09}, but we use an exact Markov chain analysis instead. Our approach gives stationary distributions of the queue lengths in addition to the mean queue lengths which can be useful for managerial decisions. Furthermore, the Markov chain approach also provides a better context for partially-collapsible state-space approach used for analyzing $SS\left(2\right)$.
\graphicspath{{Figures/}}

The state of the subsystem $SS\left(1\right)$ at a given time epoch forms a continuous time Markov chain defined by the tuple $\Big(\,l_{11}, l_{21}, r_{i1}\,\Big)$, where $l_{i1}$ is the number of products of type $i$, and $r_{i1}$ takes value of $S_{i1}$ or $U_{i1}$, for $i = 1, 2$, depending on if it is doing a setup for product $i$ or is processing product $i$. Note that $l_{11}$ and $l_{21}$ can take integer values greater than or equal to zero. Let $q\Big[\left(l_{11}, l_{21}, r_{i1}\right), \left(l_{11}', l_{21}', r_{i1}'\right)\Big]$ denote the transitions from the state $\left(l_{11}, l_{21}, r_{i1}\right)$ to the state $\left(l_{11}', l_{21}', r_{i1}'\right)$  for $\left(r_{i1}, r_{i1}'\right) \in \{S_{11}, S_{21}, U_{11}, U_{21}\}$. The transitions for the subsystem $SS\left(1\right)$ are summarized below in Table \ref{Table:6.1}.
{\renewcommand{\arraystretch}{1.30}
\begin{table}[H]
\centering
\caption{Transitions for the subsystem $SS\left(1\right)$.}\label{Table:6.1}
\begin{tabular}{| C{4cm} |  C{4cm} |  C{3cm} | C{3cm}|}
\hline
\textbf{From state} & \textbf{To state} & \textbf{Condition} & \textbf{Transition rate out}\\
\hline
$\left(l_{11}, l_{21}, S_{i1}\right)$ & $\left(l_{11}, l_{21}, S_{i'1}\right)$ & $l_{i1} ={} 0$ & \multirow{2}{*}{$\mu_{s_{i1}}$}\\
$\left(l_{11}, l_{21}, S_{i1}\right)$ & $\left(l_{11}, l_{21}, U_{i1}\right)$ &  $l_{i1} > 0$ & \\
\hline
$\left(l_{11}, l_{21}, U_{11}\right)$ & $\left(0, l_{21}, S_{21}\right)$ & $l_{11} ={} 1$ & \multirow{2}{*}{$\mu_{11}$}\\
$\left(l_{11}, l_{21}, U_{11}\right)$ & $\left(l_{11}-1, l_{21}, U_{11}\right)$ &  $l_{11} > 1$ & \\
\hline
$\left(l_{11}, l_{21}, U_{21}\right)$ & $\left(l_{11}, 0, S_{11}\right)$ & $l_{21} ={} 1$ & \multirow{2}{*}{$\mu_{21}$}\\
$\left(l_{11}, l_{21}, U_{21}\right)$ & $\left(l_{11}, l_{21}-1, U_{21}\right)$ &  $l_{21} > 1$ & \\
\hline
$\left(l_{11}, l_{21}, S_{11}\right)$ & $\left(l_{11}+1, l_{21}, S_{11}\right)$ & \multirow{4}{*}{--} & \multirow{4}{*}{$\lambda_{1}$}\\
$\left(l_{11}, l_{21}, S_{21}\right)$ & $\left(l_{11}+1, l_{21}, S_{21}\right)$ & &\\
$\left(l_{11}, l_{21}, U_{11}\right)$ & $\left(l_{11}+1, l_{21}, U_{11}\right)$ & &\\
$\left(l_{11}, l_{21}, U_{21}\right)$ & $\left(l_{11}+1, l_{21}, U_{21}\right)$ & &\\
\hline
$\left(l_{11}, l_{21}, S_{11}\right)$ & $\left(l_{11}, l_{21}+1, S_{11}\right)$ & \multirow{4}{*}{--} & \multirow{4}{*}{$\lambda_{2}$}\\
$\left(l_{11}, l_{21}, S_{21}\right)$ & $\left(l_{11}, l_{21}+1, S_{21}\right)$ & &\\
$\left(l_{11}, l_{21}, U_{11}\right)$ & $\left(l_{11}, l_{21}+1, U_{11}\right)$ & &\\
$\left(l_{11}, l_{21}, U_{21}\right)$ & $\left(l_{11}, l_{21}+1, U_{21}\right)$ & &\\
\hline
\end{tabular}
\end{table}

Let $\pi\left(l_{11}, l_{21}, r_{i1}\right)$ be the steady-state probability of state $\left(l_{11}, l_{21}, r_{i1}\right)$. The Chapman-Kolmogorov (CK) equations for the Markov chain for subsystem $SS\left(1\right)$ to and from states $\left(l_{11}, l_{21}, S_{11}\right)$ and $\left(l_{11}, l_{21}, U_{11}\right)$ are given by Equations $\left(\ref{eqn6.1}\right) - \left(\ref{eqn6.8}\right)$.\\

\setlength{\abovedisplayskip}{0pt}
\setlength{\belowdisplayskip}{0pt}
\setlength{\abovedisplayshortskip}{0pt}
\setlength{\belowdisplayshortskip}{0pt}
\begin{align*}
\intertext{For $l_{11} ={}0, l_{21} ={}0\colon$}
\left(\lambda_{1}+\lambda_{2}+\mu_{s_{11}}\right)\pi\left(0, 0, S_{11}\right)= {}\mu_{s_{21}}\pi\left(0, 0, S_{21}\right)+\mu_{21}\pi\left(0, 1, U_{21}\right)\numberthis\label{eqn6.1}\\
\intertext{For $l_{11} >{}0, l_{21} ={}0\colon$}
\left(\lambda_{1}+\lambda_{2}+\mu_{s_{11}}\right)\pi\left(l_{11}, 0, S_{11}\right)= {}\\`
\lambda_{1}\pi\left(l_{11}-1, 0, S_{11}\right)+\mu_{s_{21}}\pi\left(l_{11}, 0, S_{21}\right)+\mu_{21}\pi\left(l_{11}, 1, U_{21}\right)\numberthis\label{eqn6.2}\\
\intertext{For $l_{11} ={}0, l_{21} >{}0\colon$}
\left(\lambda_{1}+\lambda_{2}+\mu_{s_{11}}\right)\pi\left(0, l_{21}, S_{11}\right)= {}\lambda_{2}\pi\left(0, l_{21}-1, S_{11}\right)\numberthis\label{eqn6.3}\\
\intertext{For $l_{11} >{}0, l_{21} >{}0\colon$}
\left(\lambda_{1}+\lambda_{2}+\mu_{s_{11}}\right)\pi\left(l_{11}, l_{21}, S_{11}\right)= {}\lambda_{1}\pi\left(l_{11}-1, l_{21}, S_{11}\right)+\lambda_{2}\pi\left(l_{11}, l_{21}-1, S_{11}\right)\numberthis\label{eqn6.4}\\
\intertext{For $l_{11} ={}1, l_{21} ={}0\colon$}
\left(\lambda_{1}+\lambda_{2}+\mu_{11}\right)\pi\left(1, 0, U_{11}\right)= {}\mu_{s_{11}}\pi\left(1, 0, S_{11}\right)+\mu_{11}\pi\left(2, 0, S_{21}\right)\numberthis\label{eqn6.5}\\
\intertext{For $l_{11} ={}1, l_{21} >{}0\colon$}
\left(\lambda_{1}+\lambda_{2}+\mu_{11}\right)\pi\left(1, l_{21}, U_{11}\right)= {}\\
\lambda_{2}\left(1, l_{21}-1, U_{11}\right)+\mu_{s_{11}}\pi\left(1, l_{21}, S_{11}\right)+\mu_{11}\pi\left(2, l_{21}, U_{11}\right)\numberthis\label{eqn6.6}\\
\intertext{For $l_{11} >{}1, l_{21} ={}0\colon$}
\left(\lambda_{1}+\lambda_{2}+\mu_{11}\right)\pi\left(l_{11}, 0, U_{11}\right)= {}\\
\lambda_{1}\left(l_{11}-1, 0, S_{11}\right)+\mu_{s_{11}}\pi\left(l_{11}, 0, S_{11}\right)+\mu_{11}\pi\left(l_{11}+1, 0, U_{11}\right)\numberthis\label{eqn6.7}\\
\intertext{For $l_{11} >{}1, l_{21} >{}0\colon$}
\left(\lambda_{1}+\lambda_{2}+\mu_{11}\right)\pi\left(l_{11}, l_{21}, U_{11}\right) = {}\\
\lambda_{1}\left(l_{11}-1, l_{21}, U_{11}\right)+\lambda_{2}\left(l_{11}, l_{21}-1, U_{11}\right)+\mu_{s_{11}}\pi\left(l_{11}, l_{21}, S_{11}\right)+\\
\mu_{11}\pi\left(l_{11}+1, l_{21}, U_{11}\right)&\numberthis\label{eqn6.8}\\
\end{align*}

We can similarly write balance equations for states of the form $\left(l_{11}, l_{21}, S_{21}\right)$ and $\left(l_{11}, l_{21}, U_{21}\right)$. The normalization condition is written as$\colon$

\begin{align*}
\mathop{\sum\sum\sum}_{\substack{S_{i1}\in\{S_{11}, S_{21}\}\\
\left(l_{11}, l_{21}\right)\in\mathbb{Z}}}\pi\left(l_{11}, l_{21}, S_{i1}\right)
+\mathop{\sum\sum\sum}_{\substack{U_{i1}\in\{U_{11}, U_{21}\}\\
l_{i1}\in\mathbb{Z}^{+}, l_{i'1}\in\mathbb{Z}}}\pi\left(l_{11}, l_{21}, U_{i1}\right)= {}1\numberthis\label{eqn6.9}\\
\end{align*}

Using Equations $\left(\ref{eqn6.1}\right)-\left(\ref{eqn6.9}\right)$, we obtain the values of all steady state probabilities for subsystem $SS\left(1\right)$. Using the steady state probabilities, we obtain expressions for average throughput $TH_{i1}$, average buffer level $L_{i1}$, and average waiting time $W_{i1}$, of product type $i$,  for $i={}1, 2$ at station 1, and are given by Equation $\left(\ref{eqn6.10}\right)$, Equation $\left(\ref{eqn6.11}\right)$, and Equation $\left(\ref{eqn6.12}\right)$ respectively.\\
\begin{equation}\label{eqn6.10}
TH_{i1} = {}\mu_{i1}\mathop{\sum\sum}\limits_{l_{11}\in\mathbb{Z}^{+} \text{ } l_{21}\in\mathbb{Z}}\pi\left(l_{11}, l_{21}, U_{i1}\right)={}\lambda_{i}
\end{equation}

\begin{equation}\label{eqn6.11}
L_{i1} = {}\mathop{\sum\sum\sum}_{\substack{r\in\{{S_{11}, S_{21}, U_{11}, U_{21}\}}\\
\left(l_{11}, l_{21}\right)\in\mathbb{Z}}}
l_{i1}\cdot\pi\left(l_{11}, l_{21}, r\right)
\end{equation}

\begin{align*}
W_{i1}& = {}L_{i1}TH_{i1}^{-1}\numberthis\label{eqn6.12}
\end{align*}
\section{Analysis of Subsystem $SS\left(2\right)$}\label{Subsystem2}
Subsystem $SS\left(2\right)$ comprises of two-product two-station tandem polling queue as described in Section \ref{SystemDescription6} and shown in Figure \ref{fig:mesh6.1}. We perform a joint analysis of station 1 and station 2 by analyzing the Markov chain with state space aggregation. This combined analysis is necessary to incorporate the interdependencies between station 1 and station 2.
\subsection{Steady State Probabilities for $SS\left(2\right)$}\label{Subsystem2StateSpace}
To model the transitions in subsystem $SS\left(2\right)$, we have a partially-collapsible state-space description. In this description, we retain partial but relevant buffer level information for station 1, and complete and detailed buffer level information for station 2 at all time instances. We exploit the following two scenarios$\colon$

\begin{enumerate}[(a)]
\item When the server is performing setup for product $i$ at station 1, we do not track the buffer levels for any of the products at station 1, as no products are getting served at station 1. We note that if $l_{i1} > 0$ at the end of the setup, the server at station 1 will finish its setup with rate $\mu_{s_{i1}}$ and begin to serve product $i$, in which case, we will need to retrieve the buffer level information for product $i$ at station 1. The queue length retrieval for product $i$ is important to determine when the server will switch from serving product $i$ to perform setup for product $i'$. If $l_{i1} ={} 0$, the server will switch to perform setup for product $i'$, in which case, we again do not need the buffer length information for product $i'$ during its setup phase.\\

\item When the server is serving product $i$ at station 1, we only track the buffer level for product $i$ at station 1, to capture the increment in buffer levels of product $i$ at station 2, and to determine when the server switches from serving product $i$ to perform setup for product $i'$ at station 1.\\
\end{enumerate}

Through the use of this partially-collapsible state-space description, we are able to reduce the size of the state-space from one that could have six tuples to a combination of states with four-tuples and five-tuples. Our analysis shows that this loss in information does not significantly compromise the accuracy in estimates of performance measures.\\

Specifically, we define the state of the subsystem $SS\left(2\right)$ at time \emph{t} as a continuous time Markov chain defined using the following two types of states, depending on the activity of the server at station 1 at time $t\colon$
\begin{enumerate}[(i)]
	\item $\Big(S_{i1}, \,l_{12}, l_{22}, R_{i2}\,\Big)-$ When the server is performing setup at station 1$\colon$ In the state space, $S_{i1}$ represents setup for product type $i$ at station 1, $l_{i2}$ is the buffer level for type $i$ products at station 2, and $R_{i2}$ takes value of $S_{i2}$ or $U_{i2}$, for $i = 1, 2,$ depending on if the server at station 2 is doing a setup for product $i$, or is processing product $i$.\\
	
	\item $\Big(l_{i1}, U_{i1}, \,l_{12}, l_{22}, R_{i2}\,\Big)-$ When the server is serving products at station 1$\colon$ In the state space, $l_{i1}$ represents the buffer level of the product being served at station 1, $U_{i1}$ represents service for product type $i$ at station 1, $l_{i2}$ is the buffer level of type $i$ products at station 2, and $R_{i2}$ takes value of $S_{i2}$ or $U_{i2}$, for $i = 1, 2,$ depending on if the server at station 2 is doing a setup for product $i$, or is processing  product $i$.\\
\end{enumerate}

Next, we describe the state transitions for the subsystem $SS\left(2\right)$. We summarize all the state transitions for the subsystem $SS\left(2\right)$ in Table \ref{Table:6.2} below and provide explanation for the non-trivial state transitions $q\Big[\left(S_{i1}, l_{12}, l_{22}, S_{i2}\right),\left(l_{i1}, U_{i1}, l_{12}, l_{22}, U_{i2}\right)\Big]$ when $l_{i1} >{} 0$, and state-transitions $q\Big[\left(S_{i1}, l_{12}, l_{22}, S_{i1}\right),\left(S_{i'1}, l_{12}, l_{22}, S_{i'1}\right)\Big]$ otherwise. Let $p_{i\left(l_{i1}\right)}$ be the probability that there are $l_{i1}$ type $i$ products at station 1 after the server completes the setup for queue $i$. Thus, with probability $p_{i\left(l_{i1}\right)}$, for $l_{i1} >{} 0$,  there can be $l_{i1}$ type $i$ products in the queue at station 1 after the server completes setup for product $i$. In this case, the transition $q\Big[\left(S_{i1}, l_{12}, l_{22}, S_{i2}\right),\left(l_{i1}, U_{i1}, l_{12}, l_{22}, U_{i2}\right)\Big]$ occurs with rate $p_{i\left(l_{i1}\right)}\mu_{s_{i1}}$, and the server switches to serve product $i$ at station 1. Alternatively, with probability $p_{i\left(0\right)}$, the queue for product $i$ at station 1 can be empty after the server completes setup for product $i$.  Since the setups are state independent and there are 0 products in queue $i$, the transition $q\Big[\left(S_{i1}, l_{12}, l_{22}, S_{i1}\right),\left(S_{i'1}, l_{12}, l_{22}, S_{i'1}\right)\Big]$ occurs with rate $p_{i\left(0\right)}\mu_{s_{i1}}$. We determine the probability $p_{i\left(l_{i1}\right)}$ in the next section.\\

{\renewcommand{\arraystretch}{1.40}
\begin{table}[h!]
\centering
\caption{Transitions for the subsystem $SS\left(2\right)$.}\label{Table:6.2}
\begin{tabular}{| C{5cm} |  C{5cm} |  C{2.0cm} | C{2.0cm}|}
\hline
\textbf{From state} & \textbf{To state} & \textbf{Condition} & \textbf{Transition rate out}\\
\hline
\multicolumn{4}{|l|}{\textbf{Transitions at station 1.}}\\
\hline
$\left(l_{i1}, U_{i1}, l_{12}, l_{22}, R_{i2}\right)$ & $\left(l_{i1}+1, U_{i1}, l_{12}, l_{22}, R_{i2}\right)$ & & $\lambda_{i}$\\
\hline
$\left(S_{i1}, l_{12}, l_{22}, S_{i1}\right)$ & $\left(S_{i'1}, l_{12}, l_{22}, S_{i'1}\right)$ &  $l_{i1} ={} 0$ & $p_{i\left(0\right)}\mu_{s_{i1}}$\\
\hline
$\left(S_{i1}, l_{12}, l_{22}, S_{i2}\right)$ & $\left(l_{i1}, U_{i1}, l_{12}, l_{22}, U_{i2}\right)$ &  $l_{i1} > 0$ & $p_{i\left(l_{i1}\right)}\mu_{s_{i1}}$\\
\hline
\multicolumn{4}{|l|}{\textbf{Transitions at station 1 and station 2.}}\\
\hline
$\left(1, U_{11}, l_{12}, l_{22}, R_{i2}\right)$ & $\left(S_{21}, l_{12}+1, l_{22}, R_{i2}\right)$ & $l_{11} ={} 1$ & \multirow{2}{*}{$\mu_{11}$}\\
$\left(l_{11}, U_{11}, l_{12}, l_{22}, R_{i2}\right)$  & $\left(l_{11}-1, U_{11}, l_{12}+1, l_{22}, U_{11}\right)$ & $l_{11} > 1$ &\\
\hline
$\left(1, U_{21}, l_{12}, l_{22}, R_{i2}\right)$ & $\left(S_{11}, l_{12}, l_{22}+1, R_{i2}\right)$ & $l_{21} ={} 1$ & \multirow{2}{*}{$\mu_{21}$}\\
$\left(l_{21}, U_{21}, l_{12}, l_{22}, R_{i2}\right)$ & $\left(l_{21}-1, U_{21}, l_{12}, l_{22}+1, U_{11}\right)$ & $l_{21} > 1$  &\\
\hline
\multicolumn{4}{|l|}{\textbf{Transitions at station 2.}}\\
\hline
$\left(S_{i1}, l_{12}, l_{22}, S_{i2}\right)$ & $\left(S_{i1}, l_{12}, l_{22}, S_{i'2}\right)$ &  $l_{i2} ={} 0$ & \multirow{4}{*}{$\mu_{s_{i2}}$}\\
$\left(S_{i1}, l_{12}, l_{22}, S_{i2}\right)$ & $\left(S_{i1}, l_{12}, l_{22}, U_{i2}\right)$ &  $l_{i2} >{} 0$ &\\
$\left(l_{i1}, U_{i1}, l_{12}, l_{22}, S_{i2}\right)$  & $\left(l_{i1}, U_{i1}, l_{12}, l_{22}, S_{i'2}\right)$ & $l_{i2} ={} 0$ &\\
$\left(l_{i1}, U_{i1}, l_{12}, l_{22}, S_{i2}\right)$  & $\left(l_{i1}, U_{i1}, l_{12}, l_{22}, U_{i2}\right)$ & $l_{i2} >{} 0$ &\\
\hline
$\left(S_{i1}, l_{12}, l_{22}, U_{12}\right)$ & $\left(S_{i1}, 0, l_{22}, S_{22}\right)$ &  $l_{12} ={} 1$ & \multirow{4}{*}{$\mu_{12}$}\\
$\left(S_{i1}, l_{12}, l_{22}, U_{12}\right)$ & $\left(S_{i1}, l_{12}-1, l_{22}, U_{11}\right)$ &  $l_{11} > 1$ &\\
$\left(l_{i1}, U_{i1}, l_{12}, l_{22}, U_{12}\right)$  & $\left(l_{i1}, U_{i1}, 0, l_{22}, S_{22}\right)$ & $l_{12} ={} 1$ &\\
$\left(l_{i1}, U_{i1}, l_{12}, l_{22}, U_{12}\right)$  & $\left(l_{i1}, U_{i1}, l_{12}-1, l_{22}, U_{11}\right)$ & $l_{11} > 1$ &\\
\hline
$\left(S_{i1}, l_{12}, l_{22}, U_{22}\right)$ & $\left(S_{i1}, l_{12}, 0, S_{12}\right)$ &  $l_{22} ={} 1$ & \multirow{4}{*}{$\mu_{22}$}\\
$\left(S_{i1}, l_{12}, l_{22}, U_{22}\right)$ & $\left(S_{i1}, l_{12}, l_{22}-1, U_{22}\right)$ &  $l_{22} > 1$ &\\
$\left(l_{i1}, U_{i1}, l_{12}, l_{22}, U_{22}\right)$  & $\left(l_{i1}, U_{i1}, l_{12}, 0, S_{12}\right)$ & $l_{22} ={} 1$ &\\
$\left(l_{i1}, U_{i1}, l_{12}, l_{22}, U_{22}\right)$  & $\left(l_{i1}, U_{i1}, l_{12}, l_{22}-1, U_{22}\right)$ & $l_{22} > 1$ &\\
\hline
\end{tabular}
\end{table}

The CK equations for the Markov chain for subsystem $SS\left(2\right)$ are illustrated in Equations $\left(\ref{eqn6.13}\right) - \left(\ref{eqn6.20}\right)$.\\
\begin{align*}
\intertext{For $l_{12} \geq {}0, l_{22} ={}0\colon$}
\left(\mu_{s_{11}}+\mu_{s_{12}}\right)\pi\left(S_{11}, l_{12}, 0, S_{12}\right)= {}\\
\mu_{s_{22}}\pi\left(S_{11}, l_{12}, 0, S_{22}\right)+\mu_{22}\pi\left(S_{11}, l_{12}, 1, U_{22}\right)+p_{2\left(0\right)}\mu_{s_{21}}\pi\left(S_{21}, l_{12}, 0, S_{12}\right)\numberthis\label{eqn6.13}
\intertext{For $l_{12} \geq {}0, l_{22} >{}0\colon$}
\left(\mu_{s_{11}}+\mu_{s_{12}}\right)\pi\left(S_{11}, l_{12}, l_{22}, S_{12}\right)= {}\\
p_{2\left(0\right)}\mu_{s_{21}}\pi\left(S_{21}, l_{12}, l_{22}, S_{12}\right)+\mu_{21}\pi\left(1, U_{11}, l_{12}, l_{22}-1, S_{12}\right)\numberthis\label{eqn6.14}
\intertext{For $l_{12} \geq {}0, l_{22} ={}0\colon$}
\left(\mu_{12}+\mu_{s_{11}}\right)\pi\left(S_{11}, l_{12}, 0, U_{12}\right)= {}\\
\mu_{s_{12}}\pi\left(S_{11}, l_{12}, 0, S_{12}\right)+\mu_{12}\pi\left(S_{11}, l_{12}+1, 0, U_{12}\right)+p_{2\left(0\right)}\mu_{s_{21}}\pi\left(S_{21}, l_{12}, 0, S_{12}\right)\numberthis\label{eqn6.15}
\intertext{For $l_{12} \geq {}0, l_{22} >{}0\colon$}
\left(\mu_{12}+\mu_{s_{11}}\right)\pi\left(S_{11}, l_{12}, l_{22}, U_{12}\right)={}\\
\mu_{s_{12}}\pi\left(S_{11}, l_{12}, l_{22}, S_{12}\right)+\mu_{12}\pi\left(S_{11}, l_{12}+1, l_{22}, U_{12}\right)+\\
p_{2\left(0\right)}\mu_{s_{21}}\pi\left(S_{21}, l_{12}, l_{22}, S_{12}\right)+\mu_{21}\pi\left(1, U_{21}, l_{12}, l_{22}, U_{12}\right)\numberthis\label{eqn6.16}
\intertext{For $l_{11}={}1, l_{12} = {}0, l_{22} ={}0\colon$}
\left(\lambda_{1}+\mu_{11}+\mu_{s_{12}}\right)\pi\left(1, U_{11}, 0, 0, S_{12}\right)= {}\\
\mu_{s_{22}}\pi\left(1, U_{11}, 0, 0, S_{22}\right)+\mu_{22}\pi\left(1, U_{11}, 0, 1, U_{22}\right)+p_{1\left(1\right)}\mu_{s_{11}}\pi\left(S_{11}, 0, 0, S_{12}\right)\numberthis\label{eqn6.17}
\intertext{For $l_{11}={}1, l_{12} > {}0, l_{22} ={}0\colon$}
\left(\lambda_{1}+\mu_{11}+\mu_{s_{12}}\right)\pi\left(1, U_{11}, l_{12}, 0, S_{12}\right)=\\
\mu_{s_{22}}\pi\left(1, U_{11}, l_{12}, 0, S_{22}\right)+\mu_{22}\pi\left(1, U_{11}, l_{12}, 1, U_{22}\right)+\\
p_{1\left(1\right)}\mu_{s_{11}}\pi\left(S_{11}, l_{12}, 0, S_{12}\right)+\mu_{11}\pi\left(2, U_{11}, l_{12}-1, 1, U_{22}\right)\numberthis\label{eqn6.18}
\intertext{For $l_{11}={}1, l_{12} = {}0, l_{22} >{}0\colon$}
\left(\lambda_{1}+\mu_{11}+\mu_{s_{12}}\right)\pi\left(1, U_{11}, 0, l_{22}, S_{12}\right) = {}p_{1\left(1\right)}\mu_{s_{11}}\pi\left(S_{11}, 0, l_{22}, S_{12}\right)\numberthis\label{eqn6.19}
\intertext{For $l_{11}={}1, l_{12} > {}0, l_{22} >{}0\colon$}
\left(\lambda_{1}+\mu_{11}+\mu_{s_{12}}\right)\pi\left(1, U_{11}, l_{12}, l_{22}, S_{12}\right) = {}\\
p_{1\left(1\right)}\mu_{s_{11}}\pi\left(S_{11}, l_{12}, l_{22}, S_{12}\right)+\mu_{11}\pi\left(2, U_{11}, l_{12}-1, l_{22}, U_{22}\right)\numberthis\label{eqn6.20}\\
\end{align*}

Similarly, we can write balance equations for states $\Big(S_{i2}, \,l_{12}, l_{22}, R_{i2}\,\Big)$ and $\Big(l_{i2}, U_{i2}, \,l_{12}, l_{22}, R_{i2}\,\Big)$. The normalization condition is written as$\colon$\\
\begin{align*}
&\mathop{\sum\sum\sum}_{\substack{S_{i1}\in\{S_{11}, S_{21}\}\\
\left(l_{12}, l_{22}\right)\in\mathbb{Z}}}\pi\left(S_{i1}, l_{12}, l_{22}, S_{12}\right)
+\mathop{\sum\sum\sum}_{\substack{U_{i1}\in\{U_{11}, U_{21}\}\\
\left(l_{i1}, l_{12}\right)\in\mathbb{Z}^{+}, l_{22}\in\mathbb{Z}}}\pi\left(l_{i1}, U_{i1}, l_{12}, l_{22}, U_{12}\right)\\
+&\mathop{\sum\sum\sum}_{\substack{S_{i1}\in\{S_{11}, S_{21}\}\\
\left(l_{12}, l_{22}\right)\in\mathbb{Z}}}\pi\left(S_{i1}, l_{12}, l_{22}, S_{22}\right)
+\mathop{\sum\sum\sum}_{\substack{U_{i1}\in\{U_{11}, U_{21}\}\\
\left(l_{i1}, l_{22}\right)\in\mathbb{Z}^{+}, l_{12}\in\mathbb{Z}}}\pi\left(l_{i1}, U_{i1}, l_{12}, l_{22}, U_{22}\right) = {}1\\\numberthis\label{eqn6.21}
\end{align*}

Using Equations $\left(\ref{eqn6.13}\right)-\left(\ref{eqn6.21}\right)$, we obtain the estimates of all steady state probabilities for subsystem $SS\left(2\right)$. Using the steady state probabilities, we obtain estimates of the average throughput $TH_{i2}$, average buffer level $L_{i2}$, average waiting time $W_{i2}$, and system waiting time $W_{i}$, of product type $i$, for $i={}1, 2$ at station 2, these are given by Equations $\left(\ref{eqn6.22}\right)-\left(\ref{eqn6.25}\right)$.\\
\begin{align*}
TH_{i2} = {}\mu_{i2}\Big[\mathop{\sum\sum\sum}_{\substack{\left(l_{12}, l_{22}\right)\in\mathbb{Z}^{+}\\
S_{i1}\in\{S_{11}, S_{21}\}}}
\pi\left(S_{i1}, l_{12}, l_{22}, U_{i2}\right)
&+\mathop{\sum\sum\sum}_{\substack{\left(l_{11}, l_{12}, l_{22}\right)\in\mathbb{Z}^{+}}}
\pi\left(l_{11}, U_{11}, l_{12}, l_{22}, U_{i2}\right)\\
&+\mathop{\sum\sum\sum}_{\substack{\left(l_{21}, l_{12}, l_{22}\right)\in\mathbb{Z}^{+}}}
\pi\left(l_{21}, U_{21}, l_{12}, l_{22}, U_{i2}\right)\Big]={}\lambda_{i}\numberthis\label{eqn6.22}
\end{align*}
\begin{align*}
L_{i2} = {}\mathop{\sum\sum\sum\sum}_{\substack{r\in\{S_{12}, S_{22}, U_{12}, U_{22}\}\\
\left(l_{12}, l_{22}\right)\in\mathbb{Z}^{+}\\
S_{i1}\in\{S_{11}, S_{21}\}
}}
l_{i2}\cdot\pi\left(S_{i1}, l_{12}, l_{22}, r\right)
&+\mathop{\sum\sum\sum\sum}_{\substack{r\in\{S_{12}, S_{22}, U_{12}, U_{22}\}\\
\left(l_{11}, l_{12}, l_{22}\right)\in\mathbb{Z}^{+}}}
l_{i2}\cdot\pi\left(l_{11}, U_{11}, l_{12}, l_{22}, r\right)\\
&+\mathop{\sum\sum\sum\sum}_{\substack{r\in\{S_{12}, S_{22}, U_{12}, U_{22}\}\\
\left(l_{21}, l_{12}, l_{22}\right)\in\mathbb{Z}^{+}}}
l_{i2}\cdot\pi\left(l_{21}, U_{21}, l_{12}, l_{22}, r\right)\numberthis\label{eqn6.23}
\end{align*}
\begin{align*}
W_{i2}& = {}L_{i2}TH_{i2}^{-1}\numberthis\label{eqn6.24}\\
W_{i}  & = {}W_{i1}+W_{i2}, & i ={} 1, 2. \numberthis\label{eqn6.25}
\end{align*}
\subsection{Determination of $p_{i\left(l_{i1}\right)}$}\label{DeterminationOfProbability}
Next, we explain how we determine $p_{i\left(l_{i1}\right)}$. We know that $H_{ij}$ is the setup time for product $i$ at station $j$. Let $H_{j}$ be the sum of setup times for product 1 and 2 at station $j$, i.e., $H_{j} ={}H_{1j}+H_{2j}$. Further, let $V_{ij}$ denote the visit period of queue $i$, the time the server spends serving products at queue $i$ excluding
setup time at station $j$. We define intervisit period $I_{ij}$ of queue $i$ at station $j$ as the time between a departure epoch of the server from queue $i$ and its subsequent arrival to this queue at station $j$. $I_{1j}$ and $I_{2j}$ can be written as

\begin{align*}\label{eq:6.26}
I_{1j} = {} H_{2j}+V_{2j}+H_{1j}\\
I_{2j} = {} H_{1j}+V_{1j}+H_{2j}\numberthis\\
\end{align*} 

Next, we define cycle length at station $j$, $C_{j}$, as the time between two successive arrivals of the server at a particular queue at station $j$. Then, the relationship between $C_{j}$, $I_{ij}$, and $V_{ij}$ can be written as Equation $\left(\ref{eq:6.27}\right)$, and is shown in Figure \ref{fig:mesh6.4}.
\begin{equation}\label{eq:6.27}
C_{j}=H_{1j}+V_{1j}+H_{2j}+V_{2j}
\end{equation}
\graphicspath {{Figures/}}
\begin{figure}[h!]
\begin{center}
\includegraphics[scale=0.45]{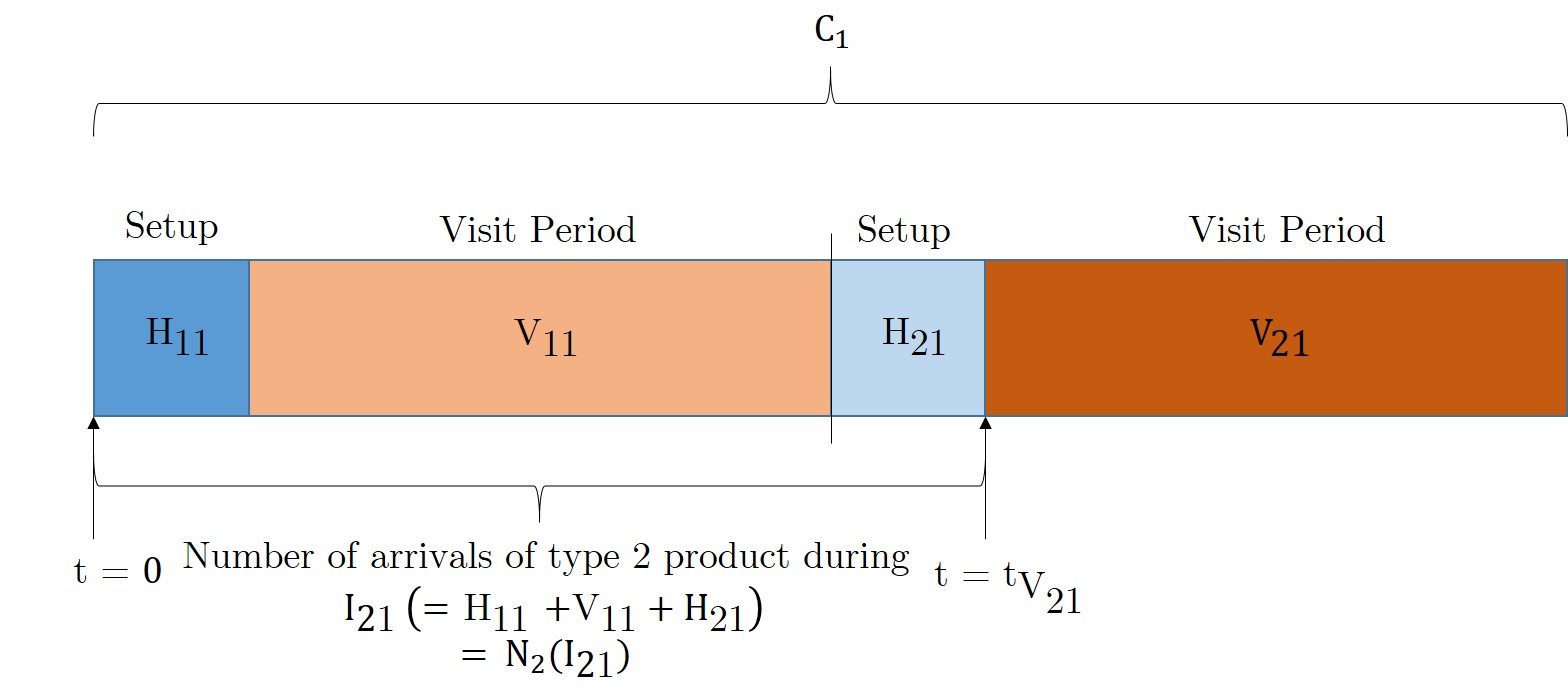}
\caption{Depiction of intervisit period $I_{21}$.}
\label{fig:mesh6.4}
\end{center}
\end{figure}

We know that $p_{i\left(l_{i1}\right)}$ is the probability that there are $l_{i1}$ type $i$ products at station 1 after the server completes the setup for queue $i$. Since the stations follow an exhaustive service policy, to calculate $p_{i\left(l_{i1}\right)}$, we need to determine the probability of a given number of Poisson arrivals at station 1 during the time interval when the server is not serving products of type $i$ at station 1, i.e, during the intervisit time of queue $i$. Note that this intervisit period is a random variable and we approximate its probability density function (pdf) using estimates of the first and the second moments of the intervisit period by method of moments.\\

Let the first moment and the variance of the setup time for product $i$ at station $j$ be $\mu_{s_{ij}}^{-1}$ and $\sigma_{s_{ij}}$ respectively. Let $\displaystyle\mathop{\mathbb{E}}\left[H_{j}\right]$ be the sum of setup times for product 1 and 2 at station $j$. Then,\\
\begin{equation}\label{eq:6.28}
\displaystyle\mathop{\mathbb{E}}\left[H_{j}\right] = {}\mu_{s_{1j}}^{-1}+\mu_{s_{2j}}^{-1}\\
\end{equation}

Next, let the traffic intensity $\rho_{ij}$ at queue \emph{i} of station \emph{j} be defined as $\rho_{ij}={}\lambda_{i}/\mu_{ij}$, and the total traffic intensity at  station \emph{j}, $\rho_{j}$, be defined as $\rho_{j}={}\sum_{i={}1}^{2} \rho_{ij}$. Note that this traffic intensity does not include the setup times. Hence, the effective load on the station is considerably higher. The mean cycle lengths in polling queues at station $j$, $C_{j}$, is given by Equation $\left(\ref{eq:6.29}\right)$.\\
\begin{equation}\label{eq:6.29}
\displaystyle\mathop{\mathbb{E}}\left[C_{j}\right] = \frac{\displaystyle\mathop{\mathbb{E}}\left[H_{j}\right]}{1 - \rho_{j}}\\
\end{equation}

Since the server is working a fraction $\rho_{ij}$ of the time on queue $i$, the mean of a visit period of queue $i$ is given by
\begin{equation}\label{eq:6.30}
\displaystyle\mathop{\mathbb{E}}\left[V_{ij}\right]={}\rho_{ij}\displaystyle\mathop{\mathbb{E}}\left[C_{j}\right]\\
\end{equation}

Therefore, the mean of intervisit period, $\displaystyle\mathop{\mathbb{E}}\left[I_{i1}\right]$, of queue $i$ at station 1 can be written as\\
\begin{equation}\label{eq:6.31}
\displaystyle\mathop{\mathbb{E}}\left[I_{ij}\right] = \mathbb{E}\left[C_{i}\right] - \mathbb{E}\left[V_{ij}\right]\\
\end{equation}

The variance of the intervisit period, $\sigma_{I_{i1}}^{2}$, of queue $i$ at station 1 is given by Equation $\left(\ref{eq:6.32}\right)$. This equation is based on the analysis by Eisenberg \cite{Eisenberg72}.\\
\begin{align*}
\sigma_{I_{i1}}^{2}=\sigma_{s_{i'1}}^{2}+\frac{\rho_{i'1}^{2}\left(\lambda_{i}T_{i1}^{2}C+\sigma_{s_{i'1}}^{2}\right) + \left(1-\rho_{i1}\right)^{2}\left(\lambda_{i'}T_{i'1}^{2}C+\sigma_{s_{i1}}^{2}\right)}{\left(1-\rho_{11}-\rho_{21}\right)\left(1-\rho_{11}-\rho_{21}+2\rho_{11}\rho_{21}\right)}\numberthis\label{eq:6.32}\\
\end{align*}

Next, we use information about $\mathbb{E}\left[I_{i1}\right]$ and $\sigma_{I_{i1}}^{2}$ in Equation $\left(\ref{eq:6.31}\right)$ and $\left(\ref{eq:6.32}\right)$ to approximate the pdf of $I_{i1}$ by a Gamma distribution. We choose the Gamma distribution since the intervisit period $I_{i1}$ is the sum of possibly non-identical exponential random variables, i.e., setup times of queue $i$ and queue $i'$, and visit period of queue $i'$. Recall that for random variable $\mathbb{Z}$, having Gamma distribution with scale parameter $\alpha$ and shape parameters $\beta$, the pdf is given by Equation $\left(\ref{eq:6.33}\right)$. The mean $\mathbb{E}\left[\mathbb{Z}\right]$ and the variance $Var\left[\mathbb{Z}\right]$ is given by Equation $\left(\ref{eq:6.34}\right)$ and Equation $\left(\ref{eq:6.35}\right)$ respectively.\\
\begin{align*}
f_{I_{i1}}\left(t\right)&={}\frac{1}{\Gamma\left(\alpha\right)\beta^\alpha}t^{\alpha-1}e^{\left(-\frac{t}{\beta}\right)}\numberthis\label{eq:6.33}\\
\mathbb{E}\left[\mathbb{Z}\right]&={} \alpha\beta\numberthis\label{eq:6.34}\\
Var\left[\mathbb{Z}\right] &={} \alpha\beta^{2}\numberthis\label{eq:6.35}\\
\end{align*}

Finally, using $f_{I_{i1}}\left(t\right)$, we determine $p_{i\left(l_{i1}\right)}$, i.e., the probability that there are $l_{i1}$ type $i$ products after the server completes the setup for queue $i$ at station 1. Let $\mathbb{N}_{i}\left(t\right)$ be the number of arrivals of product $i$ at station 1 in time $t$. Since the service policy is exhaustive at both the stations, the number of products of type $i$ at the end of the service of queue $i$ is 0 at the corresponding station. Thus, the number of type $i$ products at the end of setup for queue $i$ at station 1 is equal to the number of exogenous arrivals of type $i$ products at station 1 during the intervisit period of queue $i$. Let $l_{i1}$ be the number of type $i$ products that arrive at station 1 during the intervisit period $I_{i1}$. As the arrivals of exogenous products at station 1 are Poisson, we estimate $p_{i\left(l_{i1}\right)}$ using Equation $\left(\ref{eq:6.36}\right)$ given below.
\begin{equation}\label{eq:6.36}
p_{i\left(l_{i1}\right)}={}\Pr\left[\mathbb{N}_{i}\left(I_{i1}\right) = {}l_{i1}\right] = {}\int_{0}^{\infty}\Pr\left[\mathbb{N}\left(I_{i1} = t\right) = {}l_{i1}\right]\times f_{I_{i1}}\left(t\right)dt
\end{equation}
\section{Numerical Results}\label{NumericalResults6}
In this section, we present the results of the numerical experiments performed using the decomposition approach described in Section \ref{Subsystem1} and Section \ref{Subsystem2}. To study the accuracy of our proposed decomposition approach, a simulation model was made using Arena software (\href{https://www.arenasimulation.com/}{www.arenasimulation.com}). In the simulation model, the stations were modeled as `process' with `seize delay release' as action and products as `entities'. When the products of a particular type are processed at a station, the products of other type were held using the `hold' process. At the same time, the 'hold' process scans the queue length and releases the products of the other type when the queue length becomes zero for the served product type. A total of 10 replications were performed with a warm-up period of 50 and replication length of 500. The replication length was set to 10 days. A total of 1 million entities were processed in this duration. The simulation ran for approximately 10 minutes for each of the experimental settings.\\

To further study how our proposed approach performs against simpler models, we compared our approach with a simple decomposition approach. This simple decomposition approach looks at the system as two independent polling station. We compare the mean waiting times obtained using the proposed decomposition approach with that obtained from the simulation model  and simple decomposition under four different  experiment settings. In the first set, we compare the results under station and product symmetry. In the second set, we compare the results under station asymmetry that arises due to differences in processing rates between stations, and in the third, we compare the results under product asymmetry that arises due to differences in processing rates between products. Finally, in the fourth set, we compare the results under both station and product asymmetry. We define Error $\left(\Delta_{W_{i}}\right)$ as $|\frac{W_{i_{S}} - W_{i_{D}}}{W_{i_{S}}}|$, where $W_{i_{S}}$ and $W_{i_{D}}$ are the mean waiting times for product $i$ obtained from simulation and the decomposition approach. As expected, throughput from the decomposition model matches with the throughput from the simulation model, and the comparison of $L_{ij}$ and $W_{ij}$ give similar insights. Therefore, we focus our attention only on insights related to $W_{ij}$ in the discussion below.
\subsection{Model Validation}
\textbf{Station and Product Symmetry}$\colon$ We set the arrival rate $\lambda_{i}$ to 1.00 for both the product types at station 1 and the setup time $\mu_{s_{ij}} = {} \{1.00, 1.50, 2.00, 5.00\}$ for both the products at both the stations. We vary the service rates $\mu_{ij}$ between 2.86 to 4.00 so that the load at station $j$, $\rho_{j}$, varies between 0.50 to 0.70 in the increments of 0.10. As mentioned in Section \ref{Subsystem2}, this load does not include the setup times. Hence, the effective load on the system is considerably higher and is always 1. We also set high values for buffer sizes so that the loss in systems throughput is less than 0.1\%. The results of this comparison are summarized in Table \ref{T:6.3}. Note that, as we analyze symmetric system under this setting, $W_{1j} = {} W_{2j}$ for $j ={}1, 2$ and $W_{1} = {} W_{2}$. We do not feel the need to compare the waiting times at station 1 in our experiments as we use an exact approach to determine it.
{\renewcommand{\arraystretch}{1.35}
\begin{table}[H]
\centering
\caption{Performance analysis of systems with product and station symmetry.}\label{T:6.3}
\begin{tabular}{|C{1.0cm}|C{1.0cm}||C{1.2cm}|C{1.2cm}|C{1.2cm}||C{1.2cm}|C{1.2cm}|C{1.2cm}||C{1.2cm}|C{1.0cm}|}
\hline
\multicolumn{10}{|c|}{$\lambda_{i} = {}1, \mu_{s_{ij}}^{-1} = {}1/1.00, \text{high setup times}$}\\
\hline
\multicolumn{2}{|c||}{\textbf{Input}} & \multicolumn{3}{c||}{\textbf{Proposed Approach}} & \multicolumn{3}{c||}{\textbf{Simulation}} & \multicolumn{2}{c|}{\textbf{Error \%}}\\
\hline
&&&&&&&&&\\[-1em]
$\mu_{ij}$ & $\rho_{ij}$ & $W_{i1}$  & $W_{i2}$ &  $W_{i}$ &  $W_{i1}$  & $W_{i2}$ & $W_{i}$ & $\Delta_{W_{i2}}$ & $\Delta_{W_{i}}$\\
\hline
4.00	&	0.5	&	2.50	&	2.57	&	5.07	&	2.50	&	2.56	&	5.06	&	-0.46	&	-0.26\\
3.33	&	0.6	&	3.00	&	2.99	&	5.99	&	3.00	&	3.02	&	6.02	&	1.03	&	0.56\\
2.86	&	0.7	&	3.83	&	3.67	&	7.50	&	3.83	&	3.73	&	7.56	&	1.53	&	0.69\\
\hline
\multicolumn{10}{|c|}{$\lambda_{i} = {}1, \mu_{s_{ij}}^{-1} = {}1/1.50, \text{high-medium setup times}$}\\
\hline
\multicolumn{2}{|c||}{\textbf{Input}} & \multicolumn{3}{c||}{\textbf{Proposed Approach}} & \multicolumn{3}{c||}{\textbf{Simulation}} & \multicolumn{2}{c|}{\textbf{Error \%}}\\
\hline
&&&&&&&&&\\[-1em]
$\mu_{ij}$ & $\rho_{ij}$ & $W_{i1}$  & $W_{i2}$ &  $W_{i}$ &  $W_{i1}$  & $W_{i2}$ & $W_{i}$ & $\Delta_{W_{i2}}$ & $\Delta_{W_{i}}$\\
\hline
1.50	&	0.5	&	1.83	&	1.87	&	3.70	&	1.83	&	1.89	&	3.72	&	1.32	&	0.67\\
1.50	&	0.6	&	2.26	&	2.24	&	4.50	&	2.26	&	2.28	&	4.54	&	1.75	&	0.88\\
1.50	&	0.7	&	2.94	&	2.82	&	5.76	&	2.94	&	2.88	&	5.82	&	2.03	&	1.01\\
\hline
\multicolumn{10}{|c|}{$\lambda_{i} = {}1, \mu_{s_{ij}}^{-1} = {}1/2.00, \text{medium-low setup times}$}\\
\hline
\multicolumn{2}{|c||}{\textbf{Input}} & \multicolumn{3}{c||}{\textbf{Proposed Approach}} & \multicolumn{3}{c||}{\textbf{Simulation}} & \multicolumn{2}{c|}{\textbf{Error \%}}\\
\hline
&&&&&&&&&\\[-1em]
$\mu_{ij}$ & $\rho_{ij}$ & $W_{i1}$  & $W_{i2}$ &  $W_{i}$ &  $W_{i1}$  & $W_{i2}$ & $W_{i}$ & $\Delta_{W_{i2}}$ & $\Delta_{W_{i}}$\\
\hline
2.00	&	0.5	&	1.50	&	1.51	&	3.01	&	1.50	&	1.54	&	3.04	&	2.08	&	1.05\\
2.00	&	0.6	&	1.88	&	1.84	&	3.72	&	1.88	&	1.89	&	3.77	&	2.54	&	1.27\\
2.00	&	0.7	&	2.49	&	2.37	&	4.86	&	2.49	&	2.45	&	4.94	&	3.18	&	1.58\\
\hline
\multicolumn{10}{|c|}{$\lambda_{i} = {}1, \mu_{s_{ij}}^{-1} = {}1/5.00, \text{low setup times}$}\\
\hline
\multicolumn{2}{|c||}{\textbf{Input}} & \multicolumn{3}{c||}{\textbf{Proposed Approach}} & \multicolumn{3}{c||}{\textbf{Simulation}} & \multicolumn{2}{c|}{\textbf{Error \%}}\\
\hline
&&&&&&&&&\\[-1em]
$\mu_{ij}$ & $\rho_{ij}$ & $W_{i1}$  & $W_{i2}$ &  $W_{i}$ &  $W_{i1}$  & $W_{i2}$ & $W_{i}$ & $\Delta_{W_{i2}}$ & $\Delta_{W_{i}}$\\
\hline
5.00	&	0.5	&	0.89	&	0.87	&	1.76	&	0.89	&	0.90	&	1.79	&	2.76	&	1.57\\
5.00	&	0.6	&	1.20	&	1.14	&	2.34	&	1.20	&	1.20	&	2.40	&	4.94	&	2.37\\
5.00	&	0.7	&	1.69	&	1.55	&	3.24	&	1.69	&	1.65	&	3.34	&	5.87	&	2.83\\
\hline
\end{tabular}
\end{table}
\normalsize
It can be noted that the error in waiting times estimate at station 2 is less than 6\% while the error in system's waiting time estimates is less than 3\%  for all the tested values of traffic intensity for symmetric systems using our proposed method.\\

In the arena model simulation for all the setup settings, we see that when we vary the traffic intensity at the stations, the waiting times at station 2 which was higher than the waiting times at station 1 for lower traffic values becomes smaller for higher traffic values. This trend is captured by our proposed approach. Further, our approach is able to classify the bottleneck station for product and station symmetry settings by capturing the synergies of a tandem polling system. Although for space reasons we do not report results from the simple decomposition in the paper, we would like to point out that $i)$ the simple decomposition approach is unable to capture this trend in waiting times, and $ii)$ the simple decomposition approach yields output the same performance measure values for both the stations as it analyzes both the stations independently.\\

\textbf{Station Asymmetry Because of Different Processing Rates}$\colon$ In this experiment setting, we analyze the impact of station asymmetry by examining the effects of upstream bottlenecks and downstream bottlenecks. To do so, we first vary the service rate $\mu_{i2}$ at station 2 from 2.86 to 4.00 while keeping the service rates $\mu_{i1}$ at station 1 for both the types of products constant at 2.50. Under these settings, $\rho_{2}$ varies between 0.50 to 0.70 in the increments of 0.10. Next, to study the effects of downstream bottlenecks, we vary the service rate $\mu_{i2}$ at station 2 of both the types of products at station 2 between 2.86 to 4.00 while keeping the service rates $\mu_{i1}$ at station 1 equal at 2.50 for both the types of products. Under these settings, $\rho_{1}$ varies between 0.50 to 0.70 in the increments of 0.10. The results of this analysis are summarized in Table \ref{T:6.4}. We set the arrival rate $\lambda_{i}$ to 1.00 for both the product types at station 1 and the setup time $\mu_{s_{ij}} = {} \{1.00, 1.50, 2.00, 5.00\}$ for both the products at both the stations. Since we have only station asymmetry, $W_{1j} = {} W_{2j}$ for $j ={}1, 2$ and $W_{1} = {} W_{2}$.

\newpage

{\renewcommand{\arraystretch}{1.40}
\begin{table}[H]
\centering
\caption{Performance analysis of systems with station asymmetry.}\label{T:6.4}
\begin{tabular}{|C{1.2cm}|C{1.2cm}||C{1.2cm}|C{1.2cm}|C{1.2cm}||C{1.2cm}|C{1.2cm}|C{1.2cm}||C{1.2cm}|C{1.2cm}|}
\hline
\multicolumn{10}{|c|}{$\lambda_{i} = {}1, \mu_{i1} = {}2.50, \rho_{1} = {}0.80, \mu_{s_{ij}}^{-1} = {}1/1.00$, high setup times,  station 1 bottleneck}\\
\hline
\multicolumn{2}{|c||}{\textbf{Input}} & \multicolumn{3}{c||}{\textbf{Proposed Approach}} & \multicolumn{3}{c||}{\textbf{Simulation}} & \multicolumn{2}{c|}{\textbf{Error \%}}\\
\hline
&&&&&&&&&\\[-1em]
$\mu_{ij}$ & $\rho_{ij}$ & $W_{i1}$  & $W_{i2}$ &  $W_{i}$ &  $W_{i1}$  & $W_{i2}$ & $W_{i}$ & $\Delta_{W_{i2}}$ & $\Delta_{W_{i}}$\\
\hline
4.00	&	0.50	&	5.50	&	2.25	&	7.75	&	5.50	&	2.25	&	7.75	&	0.02	&	0.01\\
3.33	&	0.60	&	5.50	&	2.67	&	8.17	&	5.50	&	2.68	&	8.18	&	0.29	&	0.10\\
2.86	&	0.70	&	5.50	&	3.42	&	8.92	&	5.50	&	3.45	&	8.95	&	0.92	&	0.36\\
\hline
\multicolumn{10}{|c|}{$\lambda_{i} = {}1, \mu_{i2} = {}2.50, \rho_{2} = {}0.80, \mu_{s_{ij}}^{-1} = {}1/1.00$, high setup times,  station 2 bottleneck}\\
\hline
\multicolumn{2}{|c||}{\textbf{Input}} & \multicolumn{3}{c||}{\textbf{Proposed Approach}} & \multicolumn{3}{c||}{\textbf{Simulation}} & \multicolumn{2}{c|}{\textbf{Error \%}}\\
\hline
&&&&&&&&&\\[-1em]
$\mu_{ij}$ & $\rho_{ij}$ & $W_{i1}$  & $W_{i2}$ &  $W_{i}$ &  $W_{i1}$  & $W_{i2}$ & $W_{i}$ & $\Delta_{W_{i2}}$ & $\Delta_{W_{i}}$\\
\hline
4.00	&	0.50	&	2.50	&	5.86	&	8.36	&	2.50	&	5.88	&	8.38	&	0.41	&	0.29\\
3.33	&	0.60	&	3.00	&	5.71	&	8.71	&	3.00	&	5.74	&	8.74	&	0.57	&	0.38\\
2.86	&	0.70	&	3.83	&	5.47	&	9.30	&	3.83	&	5.52	&	9.35	&	0.97	&	0.57\\
\hline
\hline
\multicolumn{10}{|c|}{$\lambda_{i} = {}1, \mu_{i1} = {}2.50, \rho_{1} = {}0.80, \mu_{s_{ij}}^{-1} = {}1/1.50$, high-medium setup times,  station 1 bottleneck}\\
\hline
\multicolumn{2}{|c||}{\textbf{Input}} & \multicolumn{3}{c||}{\textbf{Proposed Approach}} & \multicolumn{3}{c||}{\textbf{Simulation}} & \multicolumn{2}{c|}{\textbf{Error \%}}\\
\hline
&&&&&&&&&\\[-1em]
$\mu_{ij}$ & $\rho_{ij}$ & $W_{i1}$  & $W_{i2}$ &  $W_{i}$ &  $W_{i1}$  & $W_{i2}$ & $W_{i}$ & $\Delta_{W_{i2}}$ & $\Delta_{W_{i}}$\\
\hline
4.00	&	0.50	&	4.33	&	1.67	&	6.00	&	4.33	&	1.69	&	6.02	&	1.18	&	0.33\\
3.33	&	0.60	&	4.33	&	2.05	&	6.38	&	4.33	&	2.06	&	6.39	&	0.49	&	0.16\\
2.86	&	0.70	&	4.33	&	2.67	&	7.00	&	4.33	&	2.72	&	7.05	&	1.84	&	0.71\\
\hline
\multicolumn{10}{|c|}{$\lambda_{i} = {}1, \mu_{i2} = {}2.50, \rho_{2} = {}0.80, \mu_{s_{ij}}^{-1} = {}1/1.50$, high-medium setup times,  station 2 bottleneck}\\
\hline
\multicolumn{2}{|c||}{\textbf{Input}} & \multicolumn{3}{c||}{\textbf{Proposed Approach}} & \multicolumn{3}{c||}{\textbf{Simulation}} & \multicolumn{2}{c|}{\textbf{Error \%}}\\
\hline
&&&&&&&&&\\[-1em]
$\mu_{ij}$ & $\rho_{ij}$ & $W_{i1}$  & $W_{i2}$ &  $W_{i}$ &  $W_{i1}$  & $W_{i2}$ & $W_{i}$ & $\Delta_{W_{i2}}$ & $\Delta_{W_{i}}$\\
\hline
4.00	&	0.50	&	1.83	&	4.53	&	6.36	&	1.83	&	4.56	&	6.39	&	0.66	&	0.47\\
3.33	&	0.60	&	2.25	&	4.41	&	6.66	&	2.25	&	4.47	&	6.72	&	1.34	&	0.89\\
2.86	&	0.70	&	2.93	&	4.19	&	7.12	&	2.93	&	4.30	&	7.23	&	2.56	&	1.52\\
\hline
\end{tabular}
\end{table}
\normalsize

{\renewcommand{\arraystretch}{1.40}
\begin{table}[H]
\ContinuedFloat  %% <-- new
\centering
\caption{Performance analysis of systems with station asymmetry (continued).}\label{T:6.4}
\begin{tabular}{|C{1.2cm}|C{1.2cm}||C{1.2cm}|C{1.2cm}|C{1.2cm}||C{1.2cm}|C{1.2cm}|C{1.2cm}||C{1.2cm}|C{1.2cm}|}
\hline
\multicolumn{10}{|c|}{$\lambda_{i} = {}1, \mu_{i1} = {}2.50, \rho_{1} = {}0.80, \mu_{s_{ij}}^{-1} = {}1/2.00$, medium-low setup times, station 1 bottleneck}\\
\hline
\multicolumn{2}{|c||}{\textbf{Input}} & \multicolumn{3}{c||}{\textbf{Proposed Approach}} & \multicolumn{3}{c||}{\textbf{Simulation}} & \multicolumn{2}{c|}{\textbf{Error \%}}\\
\hline
&&&&&&&&&\\[-1em]
$\mu_{ij}$ & $\rho_{ij}$ & $W_{i1}$  & $W_{i2}$ &  $W_{i}$ &  $W_{i1}$  & $W_{i2}$ & $W_{i}$ & $\Delta_{W_{i2}}$ & $\Delta_{W_{i}}$\\
\hline
4.00	&	0.50	&	3.75	&	1.36	&	5.11	&	3.75	&	1.40	&	5.15	&	3.07	&	0.83\\
3.33	&	0.60	&	3.75	&	1.68	&	5.43	&	3.75	&	1.74	&	5.49	&	3.51	&	1.11\\
2.86	&	0.70	&	3.75	&	2.19	&	5.94	&	3.75	&	2.32	&	6.07	&	5.52	&	2.11\\
\hline
\multicolumn{10}{|c|}{$\lambda_{i} = {}1, \mu_{i2} = {}2.50, \rho_{2} = {}0.80, \mu_{s_{ij}}^{-1} = {}1/2.00$, medium-low setup times, station 2 bottleneck}\\
\hline
\multicolumn{2}{|c||}{\textbf{Input}} & \multicolumn{3}{c||}{\textbf{Proposed Approach}} & \multicolumn{3}{c||}{\textbf{Simulation}} & \multicolumn{2}{c|}{\textbf{Error \%}}\\
\hline
&&&&&&&&&\\[-1em]
$\mu_{ij}$ & $\rho_{ij}$ & $W_{i1}$  & $W_{i2}$ &  $W_{i}$ &  $W_{i1}$  & $W_{i2}$ & $W_{i}$ & $\Delta_{W_{i2}}$ & $\Delta_{W_{i}}$\\
\hline
4.00	&	0.50	&	1.50	&	3.80	&	5.30	&	1.50	&	3.88	&	5.38	&	2.14	&	1.54\\
3.33	&	0.60	&	1.88	&	3.67	&	5.55	&	1.88	&	3.84	&	5.72	&	4.45	&	2.99\\
2.86	&	0.70	&	2.50	&	3.47	&	5.97	&	2.50	&	3.73	&	6.23	&	6.89	&	4.13\\
\hline
\hline
\multicolumn{10}{|c|}{$\lambda_{i} = {}1, \mu_{i1} = {}2.50, \rho_{1} = {}0.80, \mu_{s_{ij}}^{-1} = {}1/5.00$, low setup times, station 1 bottleneck}\\
\hline
\multicolumn{2}{|c||}{\textbf{Input}} & \multicolumn{3}{c||}{\textbf{Proposed Approach}} & \multicolumn{3}{c||}{\textbf{Simulation}} & \multicolumn{2}{c|}{\textbf{Error \%}}\\
\hline
&&&&&&&&&\\[-1em]
$\mu_{ij}$ & $\rho_{ij}$ & $W_{i1}$  & $W_{i2}$ &  $W_{i}$ &  $W_{i1}$  & $W_{i2}$ & $W_{i}$ & $\Delta_{W_{i2}}$ & $\Delta_{W_{i}}$\\
\hline
4.00	&	0.50	&	2.70	&	0.82	&	3.52	&	2.70	&	0.85	&	3.55	&	2.98	&	0.71\\
3.33	&	0.60	&	2.70	&	1.08	&	3.78	&	2.70	&	1.13	&	3.83	&	4.05	&	1.20\\
2.86	&	0.70	&	2.70	&	1.51	&	4.21	&	2.70	&	1.60	&	4.30	&	5.50	&	2.05\\
\hline
\multicolumn{10}{|c|}{$\lambda_{i} = {}1, \mu_{i2} = {}2.50, \rho_{2} = {}0.80, \mu_{s_{ij}}^{-1} = {}1/5.00$, low setup times, station 2 bottleneck}\\
\hline
\multicolumn{2}{|c||}{\textbf{Input}} & \multicolumn{3}{c||}{\textbf{Proposed Approach}} & \multicolumn{3}{c||}{\textbf{Simulation}} & \multicolumn{2}{c|}{\textbf{Error \%}}\\
\hline
&&&&&&&&&\\[-1em]
$\mu_{ij}$ & $\rho_{ij}$ & $W_{i1}$  & $W_{i2}$ &  $W_{i}$ &  $W_{i1}$  & $W_{i2}$ & $W_{i}$ & $\Delta_{W_{i2}}$ & $\Delta_{W_{i}}$\\
\hline
4.00	&	0.50	&	0.89	&	2.51	&	3.40	&	0.89	&	2.70	&	3.59	&	7.01	&	5.27\\
3.33	&	0.60	&	1.20	&	2.46	&	3.66	&	1.20	&	2.68	&	3.88	&	8.14	&	5.62\\
2.86	&	0.70	&	1.69	&	2.39	&	4.08	&	1.69	&	2.64	&	4.33	&	9.38	&	5.72\\
\hline
\end{tabular}
\end{table}
\normalsize

Table \ref{T:6.4} shows that the error in waiting times estimate using our proposed approach is less than 3\% for high and high-medium setup time settings, and is less than 10\%  for medium-low and low setup time settings. We also observe that the error in estimation of waiting times is considerably low when we have bottleneck in upstream versus when the bottleneck is in downstream operations. One important thing to notice about the system behavior is that when system parameters such as arrival rate and setup times are kept constant, the system waiting times $W_{i}$ is higher when the downstream station is a bottleneck as compared to when the bottleneck is in upstream station.\\

The major drawback of the simple decomposition approach is its inability to distinguish between bottleneck stations. In the arena simulation model and our proposed approach, we observe that system waiting times $W_{i}$ is higher when the downstream station is a bottleneck as compared to when the bottleneck is in upstream station.\\

\textbf{Product Asymmetry Because of Different Processing Rates}$\colon$ In this experiment setting, we analyze the impact of product asymmetry. For this, we fix the service rates $\mu_{1j}$ of type 1 products at both the station and vary the service rates $\mu_{2j}$ of type 2 products such that $\mu_{1j} / \mu_{2j}$ varies between 0.40 to 0.80 in the units of 0.20. We do this for  $\mu_{1j} = 2.50$. Note that in all cases, product 2 has faster service rate at both the stations. We list the results corresponding to $\mu_{s_{ij}} = {} \{1.00, 1.50, 2.00, 5.00\}$ in Table \ref{T:6.5}.

{\renewcommand{\arraystretch}{1.40}
\begin{table}[H]
\footnotesize
\centering
\caption{Performance analysis of systems with product asymmetry.}\label{T:6.5}
\resizebox{\columnwidth}{!}{%
\begin{tabular}{|C{0.9cm}|C{0.9cm}||C{0.9cm}|C{0.9cm}|C{0.9cm}|C{0.9cm}||C{0.9cm}|C{0.9cm}|C{0.9cm}|C{0.9cm}||C{0.9cm}|C{0.9cm}|C{0.9cm}|C{0.9cm}|}
\hline
\multicolumn{14}{|c|}{$\lambda_{i} = {}1, \mu_{1j} ={} 2.50, \mu_{s_{ij}}^{-1} = {}1/1.00, \text{high setup times}$}\\
\hline
\multicolumn{2}{|c||}{\textbf{Input}} & \multicolumn{4}{c||}{\textbf{Proposed Approach}} & \multicolumn{4}{c||}{\textbf{Simulation}} & \multicolumn{4}{c|}{\textbf{Error \%}}\\
\hline
&&&&&&&&&&&&&\\[-1em]
$\mu_{i2}$ & $\rho_{2}$ & $W_{12}$  & $W_{22}$ & $W_{1}$  & $W_{2}$ &  $W_{12}$ &  $W_{22}$ &  $W_{1}$  & $W_{2}$ &   $\Delta_{W_{12}}$ & $\Delta_{W_{22}}$ & $\Delta_{W_{1}}$ & $\Delta_{W_{2}}$\\
\hline
6.25	&	0.56	&	2.74	&	3.09	&	5.33	&	6.33	&	2.72	&	3.19	&	5.30	&	6.43	&	-0.90	&	3.19	&	-0.46	&	1.68\\
4.17	&	0.64	&	3.11	&	3.39	&	6.15	&	7.05	&	3.13	&	3.53	&	6.18	&	7.18	&	0.85	&	3.84	&	0.38	&	1.86\\
3.13	&	0.72	&	3.69	&	3.88	&	7.55	&	8.17	&	3.82	&	4.09	&	7.69	&	8.40	&	3.41	&	5.09	&	1.84	&	2.68\\
\hline
\multicolumn{14}{|c|}{$\lambda_{i} = {}1, \mu_{1j} ={} 2.50, \mu_{s_{ij}}^{-1} = {}1/1.50, \text{high-medium setup times}$}\\
\hline
\multicolumn{2}{|c||}{\textbf{Input}} & \multicolumn{4}{c||}{\textbf{Proposed Approach}} & \multicolumn{4}{c||}{\textbf{Simulation}} & \multicolumn{4}{c|}{\textbf{Error \%}}\\
\hline
&&&&&&&&&&&&&\\[-1em]
$\mu_{i2}$ & $\rho_{2}$ & $W_{12}$  & $W_{22}$ & $W_{1}$  & $W_{2}$ &  $W_{12}$ &  $W_{22}$ &  $W_{1}$  & $W_{2}$ &   $\Delta_{W_{12}}$ & $\Delta_{W_{22}}$ & $\Delta_{W_{1}}$ & $\Delta_{W_{2}}$\\
\hline
6.25	&	0.56	&	2.05	&	2.33	&	4.02	&	4.74	&	2.05	&	2.38	&	4.02	&	4.79	&	0.00	&	1.93	&	-0.10	&	1.02\\
4.17	&	0.64	&	2.31	&	2.58	&	4.64	&	5.34	&	2.37	&	2.66	&	4.71	&	5.43	&	2.49	&	3.01	&	1.46	&	1.66\\
3.13	&	0.72	&	2.83	&	3.00	&	5.80	&	6.29	&	2.94	&	3.14	&	5.94	&	6.46	&	3.74	&	4.49	&	2.36	&	2.65\\
\hline
\multicolumn{14}{|c|}{$\lambda_{i} = {}1, \mu_{1j} ={} 2.50, \mu_{s_{ij}}^{-1} = {}1/2.00, \text{medium-low setup times}$}\\
\hline
\multicolumn{2}{|c||}{\textbf{Input}} & \multicolumn{4}{c||}{\textbf{Proposed Approach}} & \multicolumn{4}{c||}{\textbf{Simulation}} & \multicolumn{4}{c|}{\textbf{Error \%}}\\
\hline
&&&&&&&&&&&&&\\[-1em]
$\mu_{i2}$ & $\rho_{2}$ & $W_{12}$  & $W_{22}$ & $W_{1}$  & $W_{2}$ &  $W_{12}$ &  $W_{22}$ &  $W_{1}$  & $W_{2}$ &   $\Delta_{W_{12}}$ & $\Delta_{W_{22}}$ & $\Delta_{W_{1}}$ & $\Delta_{W_{2}}$\\
\hline
6.25	&	0.56	&	1.73	&	1.86	&	3.40	&	3.86	&	1.72	&	1.98	&	3.39	&	3.98	&	-0.35	&	5.86	&	-0.18	&	3.12\\
4.17	&	0.64	&	1.93	&	2.19	&	3.91	&	4.51	&	2.00	&	2.34	&	3.98	&	4.66	&	3.40	&	6.37	&	1.71	&	3.22\\
3.13	&	0.72	&	2.39	&	2.51	&	4.94	&	5.32	&	2.51	&	2.69	&	5.07	&	5.53	&	4.90	&	6.62	&	2.66	&	3.73\\
\hline
\multicolumn{14}{|c|}{$\lambda_{i} = {}1, \mu_{1j} ={} 2.50, \mu_{s_{ij}}^{-1} = {}1/5.00, \text{low setup times}$}\\
\hline
\multicolumn{2}{|c||}{\textbf{Input}} & \multicolumn{4}{c||}{\textbf{Proposed Approach}} & \multicolumn{4}{c||}{\textbf{Simulation}} & \multicolumn{4}{c|}{\textbf{Error \%}}\\
\hline
&&&&&&&&&&&&&\\[-1em]
$\mu_{i2}$ & $\rho_{2}$ & $W_{12}$  & $W_{22}$ & $W_{1}$  & $W_{2}$ &  $W_{12}$ &  $W_{22}$ &  $W_{1}$  & $W_{2}$ &   $\Delta_{W_{12}}$ & $\Delta_{W_{22}}$ & $\Delta_{W_{1}}$ & $\Delta_{W_{2}}$\\
\hline
6.25	&	0.56	&	1.11	&	1.12	&	2.23	&	2.37	&	1.12	&	1.23	&	2.24	&	2.48	&	0.89	&	8.97	&	0.47	&	4.40\\
4.17	&	0.64	&	1.27	&	1.34	&	2.61	&	2.86	&	1.33	&	1.47	&	2.67	&	3.00	&	4.42	&	9.39	&	2.25	&	4.75\\
3.13	&	0.72	&	1.64	&	1.69	&	3.42	&	3.63	&	1.74	&	1.87	&	3.52	&	3.81	&	5.48	&	9.59	&	2.80	&	4.83\\
\hline
\end{tabular}}
\end{table}
\normalsize
Table \ref{T:6.5} shows that the error in waiting times estimate using our proposed approach is less than 4\% for high and high-medium setup time settings, and is less than 10\%  for medium-low and low setup time settings.\\

In Table \ref{T:6.5}, we observe that $W_{2}$ (for the product type having the faster service rate) is higher as compared to $W_{1}$. A possible explanation for this is that since the servers at both the stations are faster in serving products of type 2, when they switch to serve products of type 1, because of lower service rates for type 1 products, the server processes products from that queue for a longer duration. As a consequence, the products of type 2 wait longer.\\

\textbf{Station Asymmetry Because of Different Setup Rates}$\colon$ In this experiment setting, we analyze the impact of setup times on system performance. We consider the case where the upstream station is a bottleneck in terms of setup times, and set $\mu_{s_{i1}}={}1.00$ and $\mu_{s_{i1}}={}5.00$. We also consider the case where the downstream station is a bottleneck in terms of setup, and set $\mu_{s_{i1}}={}5.00$ and $\mu_{s_{i1}}={}1.00$. For both the setup settings, we vary the service rates $\mu_{ij}$ between 2.50 to 4.00 so that $\rho_{j}$ varies between 0.50 to 0.80 in the increments of 0.10. We set the arrival rate $\lambda_{i}$ to 1.00 for both the products types at station 1. The results of this analysis are summarized in Table \ref{T:6.6}.\\
{\renewcommand{\arraystretch}{1.40}
\begin{table}[H]
\centering
\caption{Performance analysis of systems with setup variation across stations.}\label{T:6.6}
\begin{tabular}{|C{1.2cm}|C{1.2cm}||C{1.2cm}|C{1.2cm}|C{1.2cm}||C{1.2cm}|C{1.2cm}|C{1.2cm}||C{1.2cm}|C{1.0cm}|}
\hline
\multicolumn{10}{|c|}{$\lambda_{i} = {}1, \mu_{s_{i1}}^{-1} = {}1/1.00, \mu_{s_{i2}}^{-1} = {}1/5.00, \text{station 1 bottleneck}$}\\
\hline
\multicolumn{2}{|c||}{\textbf{Input}} & \multicolumn{3}{c||}{\textbf{Proposed Approach}} & \multicolumn{3}{c||}{\textbf{Simulation}} & \multicolumn{2}{c|}{\textbf{Error \%}}\\
\hline
&&&&&&&&&\\[-1em]
$\mu_{ij}$ & $\rho_{ij}$ & $W_{i1}$  & $W_{i2}$ &  $W_{i}$ &  $W_{i1}$  & $W_{i2}$ & $W_{i}$ & $\Delta_{W_{i2}}$ & $\Delta_{W_{i}}$\\
\hline
4.00	&	0.50	&	2.50	&	1.00	&	3.50	&	2.50	&	0.98	&	3.48	&	-2.19	&	-0.62	\\
3.33	&	0.60	&	3.00	&	1.30	&	4.30	&	3.00	&	1.27	&	4.27	&	-2.12	&	-0.63	\\
2.86	&	0.70	&	3.82	&	1.75	&	5.57	&	3.82	&	1.72	&	5.54	&	-1.74	&	-0.54	\\
2.50	&	0.80	&	5.52	&	2.59	&	8.11	&	5.52	&	2.63	&	8.15	&	1.52	&	0.49	\\
\hline
\hline
\multicolumn{10}{|c|}{$\lambda_{i} = {}1, \mu_{s_{i1}}^{-1} = {}1/5.00, \mu_{s_{i2}}^{-1} = {}1/1.00, \text{station 2 bottleneck}$}\\
\hline
\multicolumn{2}{|c||}{\textbf{Input}} & \multicolumn{3}{c||}{\textbf{Proposed Approach}} & \multicolumn{3}{c||}{\textbf{Simulation}} & \multicolumn{2}{c|}{\textbf{Error \%}}\\
\hline
&&&&&&&&&\\[-1em]
$\mu_{ij}$ & $\rho_{ij}$ & $W_{i1}$  & $W_{i2}$ &  $W_{i}$ &  $W_{i1}$  & $W_{i2}$ & $W_{i}$ & $\Delta_{W_{i2}}$ & $\Delta_{W_{i}}$\\
\hline
4.00	&	0.50	&	0.90	&	2.43	&	3.33	&	0.90	&	2.51	&	3.41	&	3.19	&	2.35	\\
3.33	&	0.60	&	1.20	&	2.85	&	4.05	&	1.20	&	3.02	&	4.22	&	5.63	&	4.03	\\
2.86	&	0.70	&	1.69	&	3.50	&	5.19	&	1.69	&	3.79	&	5.48	&	7.65	&	5.29	\\
2.50	&	0.80	&	2.70	&	4.86	&	7.56	&	2.70	&	5.33	&	8.03	&	8.82	&	5.85	\\
\hline
\end{tabular}
\end{table}
\normalsize

Table \ref{T:6.6} shows that the error in waiting times estimate using our proposed approach is less than 3\% when we have bottleneck at upstream stations, and is less than 10\%  when we have bottleneck at downstream stations. The error values and rates show similar trend when we had station asymmetry because of different processing rates in Table \ref{T:6.4} .\\

Note that when system parameters such as arrival rate and setup times are kept constant, the system waiting times $W_{i}$ is higher when the upstream station is bottleneck in terms of setup times as compared to when the downstream station is bottleneck, when the other systems parameters are kept constant. This is opposite to the results that we observed in Table \ref{T:6.4}, when the stations where bottleneck with respect to processing times.\\

\textbf{Product Asymmetry Because of Different Setup Rates}$\colon$ Last, we compare the system performance under the settings of product asymmetry in terms of setup times. For this, we consider two settings of service rates$\colon\mu_{ij}={}2.50$ and $\mu_{ij}={}4.00$. For each of the two settings, we fix the setup rates $\mu_{s_{1j}}$ of type 1 products at both the station and vary the setup rates $\mu_{s_{2j}}$ of type 2 products such that $\mu_{s_{1j}} / \mu_{s_{2j}}$ varies between 0.40 to 0.80 in the units of 0.20. Note that in all cases, product 2 has faster setup rate at both the stations. We list the results corresponding to $\mu_{ij} = {} 2.50$ and $\mu_{ij} = {} 4.00$ in Table \ref{T:6.7}.\\
{\renewcommand{\arraystretch}{1.40}
\begin{table}[H]
\footnotesize
\centering
\caption{Performance analysis of systems with setup variation across products.}\label{T:6.7}
\resizebox{\columnwidth}{!}{%
\begin{tabular}{|C{0.9cm}|C{0.9cm}||C{0.9cm}|C{0.9cm}|C{0.9cm}|C{0.9cm}||C{0.9cm}|C{0.9cm}|C{0.9cm}|C{0.9cm}||C{0.9cm}|C{0.9cm}|C{0.9cm}|C{0.9cm}|}
\hline
\multicolumn{14}{|c|}{$\lambda_{i} = {}1, \mu_{ij}^{-1} = {}1/2.50, \text{high service times}$}\\
\hline
\multicolumn{2}{|c||}{\textbf{Input}} & \multicolumn{4}{c||}{\textbf{Proposed Approach}} & \multicolumn{4}{c||}{\textbf{Simulation}} & \multicolumn{4}{c|}{\textbf{Error \%}}\\
\hline
&&&&&&&&&&&&&\\[-1em]
$\mu_{s_{1j}}$ & $\mu_{s_{2j}}$ & $W_{12}$  & $W_{22}$ & $W_{1}$  & $W_{2}$ &  $W_{12}$ &  $W_{22}$ &  $W_{1}$  & $W_{2}$ &   $\Delta_{W_{12}}$ & $\Delta_{W_{22}}$ & $\Delta_{W_{1}}$ & $\Delta_{W_{2}}$\\
\hline
1.00	&	2.50	&	3.92	&	3.96	&	8.40	&	8.54	&	4.20	&	4.27	&	8.68	&	8.85	&	6.67	&	7.26	&	3.23	&	3.50	\\
1.00	&	1.67	&	4.15	&	4.18	&	8.92	&	9.04	&	4.49	&	4.53	&	9.26	&	9.38	&	7.57	&	7.73	&	3.70	&	3.62	\\
1.00	&	1.25	&	4.60	&	4.62	&	9.75	&	9.78	&	4.82	&	4.82	&	9.97	&	9.99	&	4.56	&	4.15	&	2.21	&	2.00	\\
\hline
\multicolumn{14}{|c|}{$\lambda_{i} = {}1, \mu_{ij}^{-1} = {}1/4.00, \text{low service times}$ }\\
\hline
\multicolumn{2}{|c||}{\textbf{Input}} & \multicolumn{4}{c||}{\textbf{Proposed Approach}} & \multicolumn{4}{c||}{\textbf{Simulation}} & \multicolumn{4}{c|}{\textbf{Error \%}}\\
\hline
&&&&&&&&&&&&&\\[-1em]
$\mu_{s_{1j}}$ & $\mu_{s_{2j}}$  & $W_{12}$  & $W_{22}$ & $W_{1}$  & $W_{2}$ &  $W_{12}$ &  $W_{22}$ &  $W_{1}$  & $W_{2}$ &   $\Delta_{W_{12}}$ & $\Delta_{W_{22}}$ & $\Delta_{W_{1}}$ & $\Delta_{W_{2}}$\\
\hline
1.00	&	2.50	&	1.97	&	2.06	&	3.87	&	4.08	&	1.95	&	2.08	&	3.85	&	4.10	&	-1.17	&	0.83	&	-0.52	&	0.49	\\
1.00	&	1.67	&	2.15	&	2.22	&	4.24	&	4.39	&	2.13	&	2.22	&	4.22	&	4.39	&	-0.81	&	0.11	&	-0.47	&	0.00	\\
1.00	&	1.25	&	2.36	&	2.39	&	4.65	&	4.71	&	2.34	&	2.38	&	4.63	&	4.70	&	-0.79	&	-0.40	&	-0.43	&	-0.21	\\
\hline
\end{tabular}}
\end{table}
\normalsize

Table \ref{T:6.7} shows that the error in waiting times estimate using our proposed approach is less than 8\% for high and high service time settings, and is less than 2\%  for low service time settings.\\

In Table \ref{T:6.7}, we observe that $W_{2}$ (for the product type having the faster setup rate) is higher as compared to $W_{1}$. A possible explanation for this is that since the servers at both the stations are faster in performing setups for products of type 2, when they switch to setup and serve products of type 1, because of lower setup rates for type 1 products, the server processes products from that queue for a longer duration. As a consequence, the products of type 2 wait longer. This observation is similar to the observed behavior of the system in Table \ref{T:6.5}.\\

Table \ref{T:6.8} summarizes the performance of the decomposition approach showing the average errors, standard deviations, and quantiles for the error \% $\left(\Delta_{W_{i2}}\right)$ and error \% $\left(\Delta_{W_{i}}\right)$. Overall, we find that the average error is around 4\% for $W_{i2}$, and around 2\% for $W_{i}$, while the errors for the majority of the cases is less than 6\%. We believe that these errors are in general satisfactory in view of the complexity of the system under consideration.\\

{\renewcommand{\arraystretch}{1.40}
\begin{table}[h!]
\centering
\caption{Summary of error analysis.}\label{T:6.8}
\begin{tabular}{|L{4.0cm}|C{4.0cm}|C{4.0cm}|}
\hline
&&\\[-1em]
\textbf{Statistics} & \textbf{Error \% $\left(\Delta_{W_{i2}}\right)$}  & \textbf{ Error \% $\left(\Delta_{W_{i}}\right)$}\\
\hline
Average error			&	3.5	&	1.8	\\
SD error					&	3.0	&	1.7	\\
$50^{th}$ quantile		&	3.0	&	1.3	\\
$75^{th}$ quantile		&	5.3	&	2.5	\\
\hline
\end{tabular}
\end{table}
\section{Conclusions}\label{Conclusions6}
In this paper, we develop a decomposition based approach to analyze tandem network of polling queues with two-products and two-stations to determine system throughput, average buffer levels, and average waiting times. Under Markovian assumptions of arrival and service times, we obtain exact values of performance measures at station 1 and use partially-collapsible state-space approach to obtain reasonably accurate approximations for performance measures at station 2. This approach allows us to analyze the system with better computational efficiency. Numerical studies are conducted to test the accuracy of the decomposition method. Overall, the average error is around 4\% for waiting time estimates at station 2, and around 4\% in estimation of system waiting times, while the errors for the majority of the cases is less than 6\%.\\

We also investigate the effects of two different types of bottleneck in the system related to product and station asymmetry, and the systems performance are different in the two cases. In the setting with station asymmetry with respect to service rates, we notice that the system waiting times $W_{i}$ is higher when the downstream station is bottleneck as compared to when the upstream station is bottleneck. However, this is not the case when there is station asymmetry with respect to setup times. In the setting with station asymmetry with respect to setup times, we observe opposite behavior. Additionally, in both cases of product asymmetry, i.e, service rates and setup rates, we observed that $W_{2}$ (for the product type having the faster service rate) is higher as compared to $W_{1}$. A simple decomposition approach that analyzes the two polling stations independently does not capture these interactions in between polling stations and gives inferior estimates of performance measures.\\

The analysis in this paper can be extended to analyze larger network of polling queues with multiple products by using product aggregation. The analysis can be also be used as building block for networks with more than two stations. Exploring these generalizations is part of our ongoing research.
\bibliographystyle{nonumber}

\end{document}